\documentclass{article}

\usepackage[utf8]{inputenc}
\usepackage[margin=1.25in]{geometry}
\usepackage{authblk}
\usepackage{setspace}
\usepackage{graphicx}
\graphicspath{{./figures/}}
\usepackage{subcaption}
\usepackage{amsmath}
\usepackage{booktabs}
\usepackage[table]{xcolor}   % colortbl: light row rules via \arrayrulecolor
\usepackage{hyperref}
\usepackage{tabularx}
\usepackage{listings}
\newcommand{\lightrule}{\arrayrulecolor{black!25}\midrule[0.4pt]\arrayrulecolor{black}}   % pale rule between table rows
\renewcommand{\topfraction}{0.9}
\renewcommand{\bottomfraction}{0.9}
\renewcommand{\textfraction}{0.1}
\renewcommand{\floatpagefraction}{0.8}

\usepackage[switch]{lineno}       % 'switch' puts line numbers on outer margins
\usepackage[
  style=nejm, 
  citestyle=numeric-comp,
  sorting=none
]{biblatex}
\title{Simulation-pretrained autoencoders versus PCA for compressing multimode fibre transmission matrices}

\author[1*]{Callum J. Wilson}
\author[1]{Terry Wright}
\author[2]{Qing Yang}
\author[2]{Zhong Wen}
\author[3]{Michael P. Pound}
\author[1]{George S.D. Gordon}

\affil[1]{OPTIMlab, Department of Chemical Engineering and Biotechnology, University of Cambridge, United Kingdom}
\affil[2]{State Key Laboratory of Modern Optical Instrumentation, Department of Optical Engineering, Zhejiang University, Hangzhou, China}
\affil[3]{School of Computer Science, University of Nottingham, United Kingdom}
\date{}

\begin{document}
%\twocolumn[
%\begin{@twocolumnfalse}

\maketitle
\let\thefootnote\relax\footnotetext{*Corresponding author: cw2011@cam.ac.uk}

%%%%%% Abstract (two-column) %%%%%%
\begin{abstract}
%The abstract should be a single paragraph written in plain language that a general reader can understand. Do not include citations, figures, tables, or undefined abbreviations in the abstract. Any abbreviations that appear in the title should be defined in the abstract. The length should not exceed 250 words, to include: 
%\begin{itemize}
%    \item An opening sentence that states the question/problem addressed by the research AND
%    \item Enough background content to give context to the study AND
%    \item A brief statement of primary results AND
%    \item A short concluding sentence.
%\end{itemize}

%Keep in mind that abstracts are the primary way research is discovered in databases. Insert keywords naturally into the abstract text, using terms that a reader might actually type into a search bar.
%\end{abstract}
%\vspace{1em}

Multimode optical fibres can image through a thin, flexible probe once their transmission matrix (TM) has been measured, but a TM is large and must be re-measured whenever the fibre moves. A compact model of feasible TMs would make storing and re-estimating them cheaper, so we ask how compactly TMs can be represented, and from how few measured TMs. Here we benchmark a permutation-equivariant autoencoder, pretrained on simulated fibres and adapted to measured TMs, against principal component analysis (PCA). At 62$\times$ compression, adapting the pretrained autoencoder with just two measured TMs reconstructs the TM of a 50\,$\mu$m fibre well enough to focus light at 98\% of the ideal enhancement, and it stays accurate when the TMs are shifted by 25 pixels or zoomed by 10\%, which quickly degrades a linear basis. It is not, however, a better compressor: PCA is about ten times more accurate when fitted on all 163 real TMs (1.1\% against 11.9\% error on held-out TMs) and is already more accurate from ten. The autoencoder's value is as a prior when few or no real TMs exist: with none it reaches 29\% error, against 91\% for PCA fitted on the same simulation. Transfer to a different fibre is only partial. The most important extension, not yet tested, is to recover pixel-basis TMs from fewer measurements than a full TM, which would turn the pretrained prior from a compressor into a measurement-saving tool.
\end{abstract}
\vspace{1em}
%\end{@twocolumnfalse}
%]

%%%%%% Main Text %%%%%%

\section{Introduction}
%Your manuscript should contain all of the numbered sections specified in this template: Introduction, Materials and Methods, Results, Discussion. Please do not reorder or retitle the section headings.
%The manuscript should start with a brief introduction that lays out the problem addressed by the research and describes the paper’s importance. The scientific question being investigated should be described in detail. The introduction should provide sufficient background information to make the article understandable to readers in other disciplines and provide enough context to ensure that the implications of the experimental findings are clear. 

Multimode fibres (MMF) guide many spatial modes of light through a core larger than a single-mode fibre, and have been proposed as cost-effective, ultrathin (\textless 0.1\,mm diameter) endoscopes \cite{wen2023stable}, promising subcellular-resolution imaging deep inside tissue \cite{phillips2022imaging}. However, MMFs suffer image distortion from bending, movement, and temperature drift \cite{zheng2023single, gordon2024flexible, amitonova2024MMFBrain, gordon2019metasurface, gordon2019fullfield}. Provided the fibre is stable, this behaviour is characterised by its transmission matrix (TM), such that $E_{\text{out}} = TM \cdot E_{\text{in}}$; knowing the TM allows these distortions to be corrected \cite{popoff2010imagetransmission, kim2015transmission}. TMs are large objects: captured experimentally in the pixel basis, i.e. as the field at each camera pixel rather than in the fibre’s mode basis, a single TM in this work occupies approximately 48.7\,MB (complex64, $42{,}025 \times 145$ elements). Compressing TMs into a compact latent representation — while retaining sufficient information for downstream reconstruction — is therefore practically important for storage, transmission, and real-time TM inference \cite{zheng2025efficient-ai-tm}.

Conventional TM calibration also requires access to the distal end of the fibre \cite{popoff2010imagetransmission}, stable reference arms, and careful alignment, limiting applicability to in vivo environments \cite{ma2026rvitm-biophotonics}. However, single-ended TM calibration is an active research area and showing promise \cite{kilpatrick2026adaptive}. Measuring one TM with 145 input patterns takes approximately 1–2 minutes on our setup, excluding fibre repositioning and alignment checks, and this scales at least linearly with the number of input patterns; because the TM changes whenever the fibre is disturbed, it must be remeasured at each new fibre state. A compact representation of the family of TMs a fibre can take, built from simulation and a few calibration measurements, could instead serve as a prior from which the current TM is estimated from limited information, with prior work already suggesting this is possible when using mode basis TMs \cite{li2021compresstms, wilson2026bend}. The rationale is that fewer degrees of freedom require fewer measurements to determine the current TM. Here we test the prerequisite: how compactly TMs can be represented, and from how few real TMs.

Principal component analysis (PCA) is a linear method that is fast to fit and requires no iterative training. Each column of a pixel-basis TM is a speckle pattern, and PCA has been used to compress speckle patterns \cite{feng2026diffusion}. Because every output speckle is a linear superposition of the fibre's guided modes, all speckles lie in a subspace whose dimension is set by the number of modes and is far smaller than the number of camera pixels. A pixel-basis TM is therefore expected to be compressible, with the limit set by the mode count. Skvarenina et al. extracted a bend-resolved modal basis from speckle-field correlation matrices using successive singular value decompositions \cite{skvarenina2025bend}, exploiting redundancy between speckle fields measured within a bend state. The redundancy in our setting is of a different kind. A TM has 145 columns, fewer than the $\approx V^2/2$ modes supported by the 50\,$\mu$m-core step index fibre used here ($V = 2\pi a\,\mathrm{NA}/\lambda \approx 52$ for core radius $a = 25\,\mu$m, $\mathrm{NA} = 0.22$ and $\lambda = 660$\,nm), so the columns of a single TM are generically linearly independent and a decomposition across them removes no redundancy. Instead, each column is a vector of 42{,}025 complex pixels that lies in the fibre's shared mode subspace, and that subspace can be learned only by pooling speckles from many TMs. Autoencoders are a nonlinear counterpart to PCA: an encoder network maps the data to a low-dimensional code and a decoder network reconstructs the data from it \cite{hinton2006reducing}, so they can capture structure that a linear subspace cannot. Whether a nonlinear autoencoder can improve on this linear baseline, and in which data regimes, remains unclear.

Deep learning autoencoders are widely used for image compression \cite{wang2022compress-ae, mienye2025ae-review}, and may be suited to TMs, which are themselves speckle-pattern images when captured in the pixel basis; convolutional networks have already been used to extract spatial features from multimode-fibre speckle patterns \cite{xiong2020dlmmf}. However, Zheng et al. \cite{zheng2024attentiontmlatent} reported that convolutional layers are ill-suited to compact mode-basis TMs, whose correlations are non-local, and used a learned invertible basis transform with self-attention to obtain sparse latent representations. Our setting differs in basis and aim: we compress pixel-basis TMs, in which each column is a speckle image with local spatial structure, and compare against PCA, a dimension-reducing linear baseline.

Training such a model requires a large corpus of TMs \cite{ma2026rvitm-biophotonics} spanning the expected distribution of fibre states, but collecting experimental TMs at scale is slow and instrument-intensive, making physics-based simulation an attractive data source. A key practical advantage of simulation is that TMs can be generated and stored in the mode basis, where dimensionality reflects the number of guided modes rather than camera pixels. In this work, simulated TMs occupy only 1.19\,MB per matrix — roughly 41$\times$ smaller than their experimental counterparts — and can be converted to the pixel basis on the fly during training, avoiding any storage bottleneck.

A critical question is whether a model pretrained on simulated TMs transfers to experimental data, which carries real-world noise, calibration artefacts, and physical imperfections absent from simulation. Fibres of different lengths present an additional challenge: simulating a 5\,m fibre accurately requires substantially more compute than simulating a 1\,m fibre, raising the question of whether cheaper 1\,m simulations provide a useful pretraining signal for experimental 5\,m TMs. We therefore construct 5\,m TMs by multiplying five 1\,m simulated TMs (``composite’’ TMs), and ask whether these provide a useful pretraining signal.

Here, we compare a simulation-pretrained autoencoder with PCA on pixel-basis TMs of a 5\,m multimode fibre, and ask seven questions:
\begin{enumerate}
    \item How redundant is a pixel-basis TM: how much can PCA compress it, and what sets the limit?
    \item Can ``composite'' simulated TMs, formed by multiplying several 1\,m simulated TMs, provide a useful pretraining signal for 5\,m experimental TMs?
    \item Does a nonlinear autoencoder compress TMs better than PCA at the same latent size?
    \item How many real TMs does each approach need, and where does a pretrained prior help?
    \item How large is the simulation-to-experiment gap, and can a pretrained model bridge it where a linear basis cannot?
    \item Do the models transfer to a different fibre, and can one model serve several? We evaluate a second, larger-core fibre.
    \item Do the remaining reconstruction errors matter for focusing and imaging, and what survives geometric drift, such as camera shift and zoom?
\end{enumerate}

The autoencoder here serves as a vehicle for studying the simulation-pretraining workflow itself: pretrain on simulated TMs, then adapt to a fibre using few or no real TMs. The questions above concern how many real TMs this workflow needs, which we expect to be relevant to other TM models, although we test only this architecture. In our data, PCA is hard to beat when real TMs are available, whereas a simulation-pretrained autoencoder is most useful when they are not.

\section{Materials and Methods}
%The materials and methods section should provide sufficient information to allow replication of the results. This section should be broken up by subheadings. Under exceptional circumstances, when a particularly lengthy description is required, a portion of the materials and methods can be included in the Supplementary Materials. 
%Begin with a section titled Experimental Design describing the objectives and design of the study as well as prespecified components.

\subsection{Simulated data}

Multiple simulation techniques exist, such as: the transfer matrix method (TMM), which models the fibre as a sequence of infinitesimally small segments and uses a spatial translation between segments to simulate the mode coupling induced by bending \cite{argyros2008tmm}; pyMMF \cite{popoff2024pymmf}; and the beam propagation method (BPM) \cite{veettikazhy2021bpm}, which numerically solves the full propagating field and, in our experience, scales poorly to the fibre lengths and segment counts required here. We adopt TMM here, as it has been validated against experimental MMF transmission matrices in prior work from our group \cite{zheng2023single, wilson2026bend}, and its computational cost scales with fibre length — a property this paper exploits directly when comparing cheap 1\,m simulations against costly 5\,m simulations. The simulated fibre had the following properties: fibre length 1\,m, numerical aperture 0.22, refractive index 1.457, wavelength 632.8\,nm (the experimental laser was 660\,nm; this mismatch was unintentional and means the simulated fibre has $V \approx 55$ against $\approx 52$ for the experiment, which we treat as part of the simulation-to-experiment gap), core diameter 50\,$\mu$m, with the number of guided modes kept being 386 (scalar LP modes with one polarisation and one azimuthal orientation). Simulated TMs were generated using randomly assigned bend radii and directions which are saved alongside the mode-basis TM and can be traced. Each generated 1\,m mode-basis TM must have a 2-norm condition number calculated using singular value decomposition (SVD) of less than 250 to remain in the dataset. 5{,}000 simulated 1\,m TMs were generated and split 80:10:10 into training, validation and test sets with a fixed seed (as for the experimental data), so no 1\,m TM appears in more than one split.

The 5\,m fibre was simulated as a composite: each 5\,m TM is the product, in the mode basis, of five 1\,m TMs drawn without replacement from the same split's pool, with the product taken in the propagation order of the fibre. The five indices of every composite were drawn once with a fixed seed and stored, so every run uses identical composites. Training sets of 500, 1{,}000 and 2{,}000 composites are nested prefixes of one list, and the validation and test sets each contain 250 composites. Composite TMs are built on the fly during training: the stored 1\,m TMs are loaded, multiplied, and converted to the pixel basis. Mode amplitudes for 145 fixed random orthonormal input patterns are propagated through the composite, and a fixed mode-to-pixel matrix maps the output to the $205 \times 205$ camera grid, giving a $42{,}025 \times 145$ pixel-basis TM. Gaussian amplitude and phase noise was then added to each TM, with the amplitude and phase noise levels drawn uniformly from 1--15\% and 0.01--0.1\,rad per TM. Each TM was normalised by its largest singular value.

\subsection{Experimental data}

Experimental data (204 TMs) was collected using a 5\,m multimode fibre (50\,$\mu$m core diameter, numerical aperture 0.22). These 204 TMs were created by randomly altering the bend state of the MMF, whose ends were fixed to an optical table. Light from a 660\,nm laser was modulated using a spatial light modulator, which displayed 145 input patterns, each generating a tilted plane wave at the fibre input. The 145 inputs are mutually orthogonal (the correlation between every pair is zero), and the number was limited by the stability of the setup over the measurement time. A single mode fibre carried the reference beam, which was interfered with the beam exiting the fibre at the distal facet to measure the phase by off-axis holography. A charge-coupled device (CCD) collected the resulting output speckle. Each element of both the simulated and experimental TMs is a complex number $a + bi$. The TMs were normalised independently by their own largest singular value. Both simulated and experimental data scaling studies use nested datasets, for example the 50 TM dataset would contain the exact same TMs in the 25 TM dataset, plus 25 extra. This ensures that the effect of more data is isolated, not the data itself. The training, validation, test sets were split 80:10:10, resulting in 163, 20, 21 TMs, respectively.

\subsection{Second-fibre experimental data}

To test transfer to a different fibre, we used 164 TMs measured by two of the authors \cite{wen2023stable} on a larger-core fibre (105\,$\mu$m core, 1{,}024 input patterns, $64 \times 64$ output pixels), split 131, 16, 17 into training, validation and test sets. These TMs are not part of our public release and can be requested from the authors who measured them (Z.W. and Q.Y.; see \cite{wen2023stable}). Of the 1{,}024 inputs we kept the 145 with the highest mean column power, averaged over the training TMs, to match the 145 inputs of the original data. The same 145 inputs were used for every TM in the training, validation and test sets. Each output field was upscaled to $205 \times 205$ by zero-padding its two-dimensional Fourier spectrum, and each TM was divided by its largest singular value, as for the original TMs. We call these the raw second-fibre TMs. A coherent speckle is band-limited: its spectrum fills a disc in spatial frequency whose radius is set by the aperture, here the fibre’s numerical aperture \cite[Ch.~6]{goodman1996fourier}, while the core diameter sets the field of view \cite{cizmar2011shapinglight}. A larger core therefore fits more speckle grains across the frame, so the second fibre’s speckle patterns carry finer spatial structure than those of the 50 µm fibre: the Fourier disc holding 99\% of a speckle’s power contains $\pi r_{99}^2 \approx 4{,}300$ complex degrees of freedom, against about 800 for the original data. The two fibres differ in many ways (core size, mode count, setup and input patterns), so to test the effect of the larger core's finer speckle on its own, we also made band-limited second-fibre TMs. Before upscaling, each output field was low-pass filtered in the Fourier domain with a disc of radius $R = 15$ cycles per frame (raised-cosine edge over two bins), chosen so that the number of modes per speckle matches the original data (Supplementary Note S1). Band-limiting discards spatial detail, so errors on band-limited and raw TMs are not directly comparable.

\subsection{Model architecture}

\begin{figure}[htbp]
    \centering
    \includegraphics[width=1\textwidth]{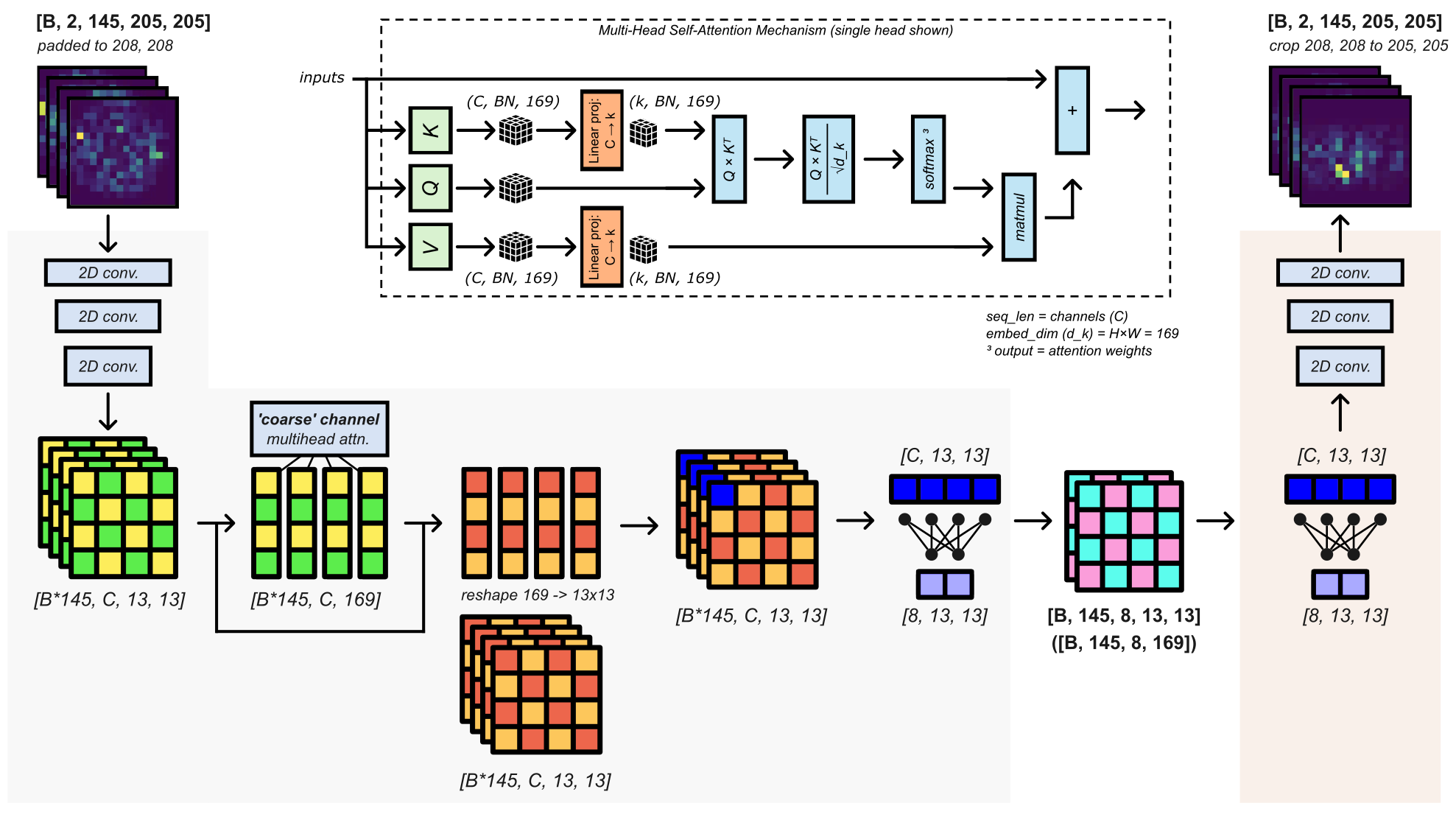}
    \caption{Autoencoder architecture. A TM of shape $[B, 2, 145, 205, 205]$ (batch, real and imaginary channels, input pattern, output image) is padded to $208 \times 208$, and each of its 145 speckle images is processed independently with shared weights. Encoder (left and bottom): three 2D convolutional layers reduce each image to $C$ feature maps of $13 \times 13$. Channel self-attention (top, one head shown) then treats the $C$ channels as the sequence and the flattened $13 \times 13 = 169$ positions as the embedding. As in the Linformer \cite{wang2020linformer}, the keys and values are linearly projected from $C$ to $k$ channels, which shortens the sequence and reduces the attention matrix from $C \times C$ to $C \times k$, and the attention output is added to its input (residual connection). A linear layer across channels then reduces $C$ to the latent channel count, giving a latent of shape $[145, 8, 13, 13]$ at 62$\times$ compression (4 channels at 124$\times$). Decoder (right): a linear layer across channels and three 2D convolutional layers restore the $208 \times 208$ image, which is cropped to $205 \times 205$. The decoder is not a mirror of the encoder and contains no attention, so decoding is fast. There are no skip connections between the encoder and decoder. Channel widths, $k$ and the number of heads are given in Methods.}
    \label{fig:autoencoder_arch}
\end{figure}

The encoder and decoder are trained without skip connections between them, ensuring the compressed latent representation is a complete and self-contained encoding of the transmission matrix (Figure~\ref{fig:autoencoder_arch}). This design choice allows the decoder to be used independently of the encoder — for instance, reconstructing a transmission matrix from a latent vector obtained by means other than the trained encoder — a property that would be precluded by skip connections directly propagating encoder activations into the decoder. The decoder is also deliberately not a mirror of the encoder: it contains no attention, only a linear layer and convolutions, which makes decoding fast when it is used on its own. Without skip connections, error may increase.

The autoencoder is permutation equivariant over the input patterns. A TM of shape $42{,}025 \times 145$ is treated as 145 two-channel (real and imaginary) $205 \times 205$ images, each processed independently with shared weights; the channel attention acts across feature channels within a single speckle, never across inputs. Reordering the columns of a TM therefore reorders the reconstruction in the same way, so the model does not depend on the order of the input basis. This matters for three reasons. First, the columns of a TM have no canonical order, since they depend on how the input patterns are chosen. Second, the same weights can be pretrained on simulated inputs and then applied to an experiment, or to another fibre, that uses a different set of input patterns: in this work a model pretrained on simulated TMs with random orthonormal input patterns is fine-tuned on, and applied to, experimental TMs measured with tilted plane waves, and to the second fibre's TMs measured with a different set of inputs. Third, each TM supplies 145 training examples, which makes better use of a small number of TMs than treating a whole TM as one sample. The cost is that correlations between different input patterns are not exploited.

The architecture follows preliminary tests on smaller $256 \times 256$ TMs, run before the experimental data were collected. In those tests Linformer-style \cite{wang2020linformer} coarse channel attention matched full channel attention after fine-tuning (45.3\% against 44.6\% experimental error) with a smaller attention matrix, and a one-dimensional convolutional encoder failed to train (Supplementary Table~\ref{tab:sm-attention_variants} and Figures~\ref{fig:sm-attn_full}--\ref{fig:sm-attn_1dconv}). A complex-valued autoencoder of the same design did not match the two-channel (real and imaginary) model on the pixel-basis TMs. At the same latent size (2{,}704 real numbers per speckle, i.e.\ 8 complex against 16 real channels) it failed to train under our recipe (101\% error after fine-tuning, against 7.5\%), and with twice as many latent numbers (16 complex channels) it trained but gave a higher fine-tuned error (13.1\%) (Supplementary Table~\ref{tab:sm-complex_vs_2channel}). These were single runs under earlier protocols, so we use them to motivate the design and not to rank architectures.

The compression rate of the autoencoder is controlled by the number of channels in the latent space, the feature map size is fixed throughout. A single TM has shape [145, 2, 205, 205] (real), some example compression rates and latent tensor shapes below:

\begin{enumerate}
    \item Compression rate 62$\times$: [145, 8, 13, 13] (real)
    \item Compression rate 124$\times$: [145, 4, 13, 13] (real)
\end{enumerate}

\textit{Note that each speckle pattern was padded to size [208, 208] to allow for more natural convolution feature reduction.}

\subsection{PCA baseline}

Each TM was treated as 145 speckle vectors, each holding the real and imaginary parts of its 42{,}025 pixels (84{,}050 reals). A single PCA was fitted over all speckles of the training TMs, after subtracting their mean, so the basis is shared across columns and across TMs. A TM was compressed by projecting each speckle onto the first $k$ components and reconstructed from those coefficients. The 62$\times$ setting uses $k = 1{,}352$, matching the autoencoder latent of $8 \times 13 \times 13 = 1{,}352$ reals; with fewer than $k + 1$ training speckles, $k$ is capped at one less than the number of speckles. PCA involves no iterative training and was fitted once for each training set, so no seeds apply. Errors are scored with Eq.~\ref{eq:tm_error}.

\subsection{Training and finetuning}

All models were trained with AdamW and gradient-norm clipping (maximum norm 1.0), minimising the phase-aligned error of Eq.~\ref{eq:tm_error}. The final configuration (encoder widths $[24, 48, 72]$, one attention head, Linformer projection length $k = 24$, pretraining learning rate $1.35 \times 10^{-3}$, pretraining batch size 4, fine-tuning learning rate $1.49 \times 10^{-4}$, weight decay $1.27 \times 10^{-6}$) was chosen by a Bayesian search (Optuna, TPE sampler) \cite{optuna_2019} over these quantities with validation error after fine-tuning as the objective. Altogether 27 trials were undertaken. Sections~\ref{original_training_recipe_starts}--~\ref{original_training_recipe_ends} of Results use the baseline configuration that preceded the search (encoder widths $[16, 32, 48]$, $k = 16$, pretraining learning rate $1.53 \times 10^{-3}$, batch size 8), which we call the baseline configuration.

Training has two stages. Unless stated otherwise, pretraining used 2,000 simulated composite TMs, with fresh noise generated for every batch and a new random order every epoch. After each epoch the model is scored on the 20 experimental validation TMs and the lowest-error epoch is kept; pretraining stops after 30 epochs without improvement (epoch cap 1{,}000). Early stopping therefore uses experimental TMs to pick the checkpoint, even for the pretrained model, so the validation errors are slightly optimistic estimates of performance on unseen TMs (the test TMs play no part in this choice). This applies equally to the fine-tuned and the experiment-only models. In fine-tuning, the best pretrained weights are trained further on $N$ experimental TMs (batch size 2), starting from the optimiser state at the end of pretraining with the learning rate reset to the fine-tuning value. The checkpoint is again the lowest error on the 20 validation TMs, with patience 30 and an epoch cap of 1{,}000. Experiment-only baselines use the same architecture and the pretraining recipe, trained from scratch on the experimental TMs without simulation.

The few-shot, second-fibre and shared-model experiments start from the same pretrained checkpoints but use a simpler fine-tuning recipe: a fresh AdamW optimiser (the end-of-pretraining optimiser state was not saved), batch size 2, patience 30 and an epoch cap of 5{,}000. A ``specialist'' is a model fine-tuned on TMs of a single fibre only; the second-fibre specialist is fine-tuned on the first $N$ of its 131 training TMs, with the checkpoint chosen on its own 16 validation TMs; the band-limited variant uses the same recipe on band-limited TMs. A shared model is fine-tuned on the first $N$ training TMs of each fibre ($2N$ in total), shuffled together every epoch, with the checkpoint chosen on the mean of the two fibres' validation errors so that each fibre counts equally. Zero-shot results use the pretrained checkpoint unchanged. PCA comparators are fitted on the same TMs, with a single basis fitted on all $2N$ TMs for shared models.

All experiments were run with multiple seeds, either three or five; five were used when the experiment was cheaper, such as training on a small experimental dataset. Scores are averaged over seeds and reported with the standard deviation.

A run is counted as ``converged’’ if its best validation error is below 90\%. In our experiments a run either trained to close to the floor for that experiment or stayed at around 100\%. Tables report the mean $\pm$ SD over converged seeds and over all seeds, with the number of converged seeds.

\subsection{Evaluation}
Reconstructions are scored with the phase-aligned relative error

\begin{equation}
\text{TM error (\%)} = \frac{ \left\| B - A \, e^{-j \, \arg \left( \operatorname{tr} \left( A B^H \right) \right)} \right\|_F }{ \|A\|_F } \times 100,
\label{eq:tm_error}
\end{equation}

where $A$ is the measured (or simulated) TM, $B$ is its reconstruction, $\|\cdot\|_F$ is the Frobenius norm and $^H$ the conjugate transpose. The phase factor is the single global phase that minimises the error for each TM, because the overall phase of a measured TM is set by an arbitrary reference and carries no information about the fibre. Errors are averaged over the TMs of a dataset.

Two further tests judge a reconstruction by what it does when used. In both, an input is computed from the reconstructed TM and applied to the true TM, and the result is compared with the same procedure run on the true TM, which is the ideal for 145 inputs. For focusing, the input is the phase conjugate of the reconstructed TM's row at a random output pixel inside the bright disc, and we report the focus enhancement (peak over mean intensity) on the true TM as a fraction of the ideal. For imaging, the input that best produces a target intensity pattern (a letter F, a set of spots or a ring) is found from the reconstructed TM with Gerchberg--Saxton iterations: each iteration solves a regularised least-squares (pseudo-inverse) problem for the input that best produces the current target field, then replaces the amplitude of the resulting output field with the target amplitude while keeping its phase. The output intensity on the true TM is compared with the target by Pearson correlation and SSIM. To test robustness to drift, the test TMs are perturbed before any model sees them (shift, zoom, rotation, blur or added noise) and each reconstruction is scored against the perturbed TM with Eq.~\ref{eq:tm_error}. A shift moves the $205 \times 205$ output image of every input by a number of pixels; pixels moved past the edge are lost and the vacated strip is filled with weak complex Gaussian noise, like a camera that has moved.
\vspace*{1em}

\subsection{Reproducibility}

Each run is identified by a seed, which fixes the weight initialisation, the order of training batches and the simulation noise generation. The Python, NumPy and PyTorch generators were seeded. The final-configuration and test pipelines used deterministic cuDNN/cuBLAS kernels with TF32 disabled; earlier pipelines set deterministic cuDNN only. Model seeds were either 42--44 or 42--46 during analysis and training, relevant values can be found in Results. The test results are single seeded (42). The data splits use a separate fixed seed (42), and the test splits were scored once, for a frozen list of models, with the model trained under seed 42 (i.e. no seed averaging). Experiments were run on the following hardware and software specifications: OS = Windows 11, GPU = Nvidia RTX Pro 6000 Blackwell Max-Q Workstation edition, Python = 3.12.14, PyTorch = 2.14.0+cu130 (CUDA 13.0 runtime).

Early experiments were run by a driver that restarted the training process for every sub-batch of 100 composite TMs, to fit in GPU memory, restoring the weights and optimiser state between sub-batches; there an epoch is one pass over all sub-batches and early stopping used the mean of the sub-batch validation errors. This applies to Sections~\ref{original_training_recipe_starts}--~\ref{original_training_recipe_ends}.

\section{Results}

The results follow the questions in the Introduction. Sections ~\ref{Compression-limits-of-pixel-basis-TMs}--~\ref{Autoencoder-vs-PCA-at-62x-on-original-experimental-data} establish how compressible pixel-basis TMs are (Q1), whether composite simulations help (Q2), and how a pretrained autoencoder compares with PCA (Q3--Q5); these design experiments use the validation set. Sections ~\ref{Collaborator-fibre:-zero-shot,-few-shot-and-shared-model}--~\ref{Shared-finetuning} test transfer to a second fibre (Q6). Section ~\ref{Final-test-scores-and-robustness-tests} reports the final comparison on held-out test TMs, where all models were frozen first, and tests robustness to geometric drift (Q7).

\subsection{Compression limits of pixel-basis TMs}
\label{Compression-limits-of-pixel-basis-TMs}

\begin{figure}[!htbp]
    \centering
    \includegraphics[width=0.65\textwidth]{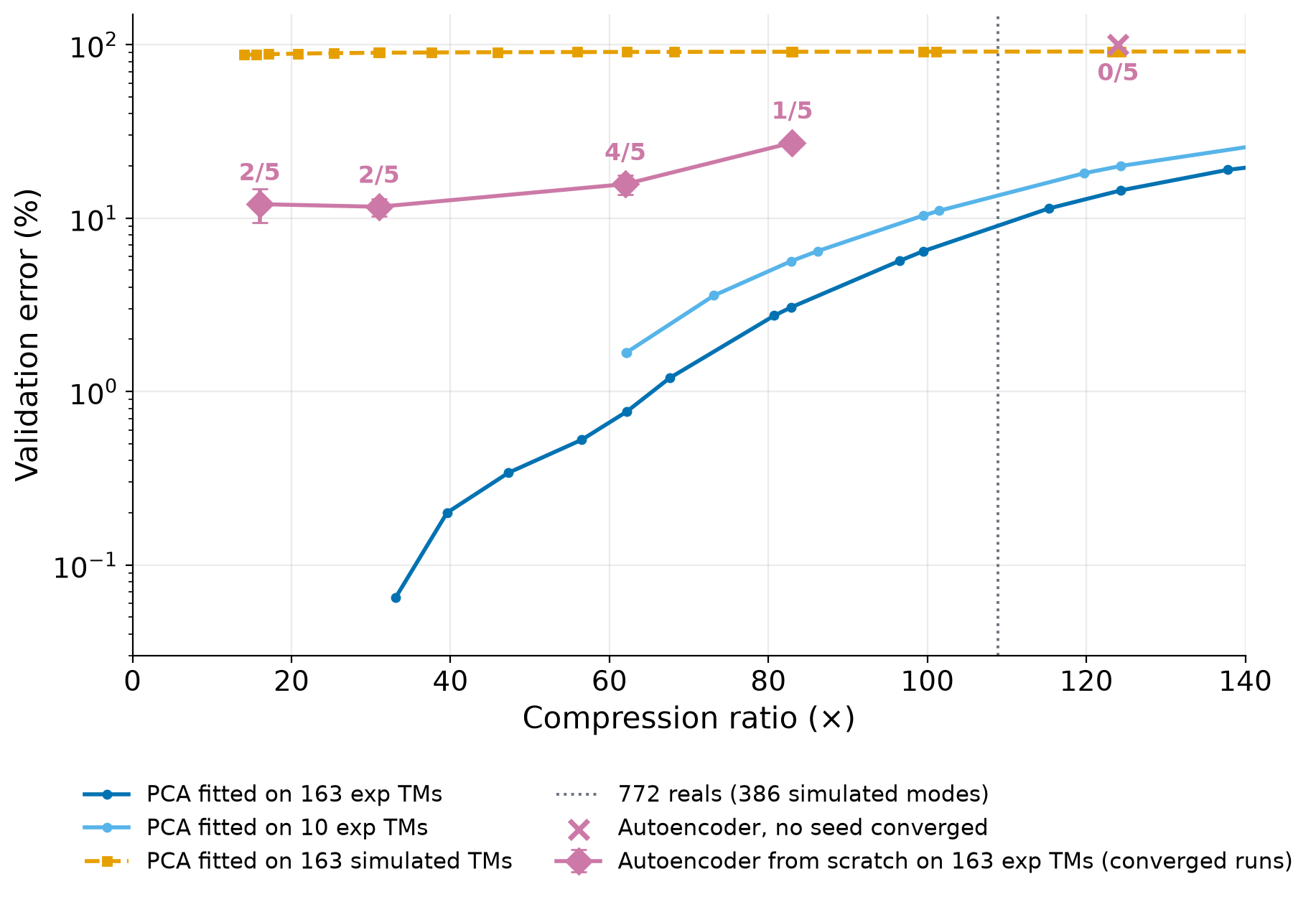}
    \caption{Validation error against compression ratio on the original experimental TMs. Lines: PCA fitted on 163 or 10 experimental TMs, or on 163 simulated TMs; the dotted vertical line marks 772 reals (the 386 simulated modes). Diamonds: autoencoder trained from scratch on the 163 experimental training TMs, mean $\pm$ SD of the converged seeds, labelled with the number of converged seeds out of 5 (the 83$\times$ point is a single run, so has no SD; at 124$\times$ no seed converged and the cross is the mean of all seeds). Only 2,538 PCA components rise above the float32 noise floor, so the PCA curves end at around 33$\times$. Per-seed autoencoder results are in Supplementary Table~\ref{tab:sm1-autoencoder_exp_compression_study}.}
    \label{fig:pca_vs_ae_compression}
\end{figure}

PCA reaches 0.77\% error at 62$\times$ (1,352 components), 3.1\% at 83$\times$ and 14.4\% at 124$\times$, and 0.07\% at 33$\times$, where the curve ends at the float32 noise floor (Figure~\ref{fig:pca_vs_ae_compression}). It needs about ten experimental TMs to fit well (1.7\%; 25.3\% with five, Table~\ref{tab:validation_error_vs_num_tms}), and a basis fitted on one domain fails on the other (90.8\% for simulation-fitted PCA scored on experiment, Table~\ref{tab:pca_domain_transfer}), which is the simulation-to-experiment gap (Q5). This answers Q1: pixel-basis TMs are highly compressible, with the limit set by the mode count.

\subsubsection{PCA}
\label{PCA}

PCA relies on having enough real TMs to span the mode subspace. Each TM supplies 145 speckles, so a fit on $N$ TMs can have at most $145N - 1$ components, and the 1,352 components of a 62$\times$ code need $N \ge 10$. The error therefore falls from 89.5\% with one TM and 25.3\% with five (724 components) to 1.7\% with ten, and then improves only slowly (1.3\% with 25 TMs, 0.8\% with 163; Table~\ref{tab:validation_error_vs_num_tms}). This is the PCA half of Q4: below five real TMs PCA is not usable at 62$\times$ but around ten are enough, so any advantage of a simulation-pretrained prior must lie below that.

\begin{table}[htbp]
\begin{center}
\begin{tabular}{cccccccc}
 \toprule
 Experimental TMs used & 1 & 2 & 3 & 5 & 10 & 25 & 163 \\
 \midrule
 62$\times$ validation error (\%) & 89.5 & 65.6 & 52.4 & 25.3 & 1.7 & 1.3 & 0.8 \\
 \bottomrule
\end{tabular}
\caption{PCA performance on experimental TMs when using varying number of TMs to fit the model.}
\label{tab:validation_error_vs_num_tms}
\end{center}
\end{table}

A PCA basis does not transfer between domains in either direction: fitted on 163 simulated composite TMs it gives 10.3\% error on held-out simulated TMs but 90.8\% on experimental TMs, and fitted on experiment it gives 0.8\% in-domain but 38.5\% on simulated TMs (Table~\ref{tab:pca_domain_transfer}). This is the simulation-to-experiment gap (Q5) that a simulation-pretrained model has to bridge.

\begin{table}[htbp]
\begin{center}
\begin{tabular}{lcc}
 \toprule
 & \multicolumn{2}{c}{Validation error (\%) [Complex correlation]} \\ \cmidrule(lr){2-3}
 Dataset used to fit PCA & Experimental & Simulated \\
 \midrule
 163 exp TMs & 0.77 [1.00] & 38.48 [0.90] \\
 163 simulated composite-5m TMs & 90.78 [0.42] & 10.28 [0.99] \\
 \bottomrule
\end{tabular}
\caption{PCA fitted on one domain does not transfer to the other as well, especially in the case of fitting PCA on simulated data and evaluating on experimental.}
\label{tab:pca_domain_transfer}
\end{center}
\end{table}

\subsubsection{Autoencoder}
\label{original_training_recipe_starts}

An autoencoder trained from scratch on experimental TMs is far worse and unreliable. At 62$\times$ only 4 of 5 seeds converged, to $15.7 \pm 2.0$\%, against 0.77\% for PCA, and no seed converged at 124$\times$ or with fewer than 150 TMs (Figure~\ref{fig:pca_vs_ae_compression} and Table~\ref{tab:autoencoder_exp_data_scaling_study}). This answers the from-scratch half of Q3: without simulation pretraining the autoencoder does not compress better than PCA.

\begin{table}[htbp]
\begin{center}
\begin{tabular}{cccc}
 \toprule
 & & \multicolumn{2}{c}{Validation error (\%)} \\ \cmidrule(lr){3-4}
 N. training TMs & Seeds converged & Converged avg. $\pm$ SD & All-seed avg. $\pm$ SD \\
 \midrule
 100 & 0/5 & n/a & $100.89 \pm 1.50$ \\
 125 & 0/5 & n/a & $100.57 \pm 0.55$ \\
 150 & 2/5 & $13.59 \pm 0.76$ & $65.65 \pm 47.53$ \\
 163 & 4/5 & $15.70 \pm 2.00$ & $32.64 \pm 37.93$ \\
 \bottomrule
\end{tabular}
\caption{Autoencoder experimental-only data scaling study, with fixed compression rate of 62$\times$ (the 163-TM row is the same runs as the 62$\times$ point of Figure~\ref{fig:pca_vs_ae_compression}). A 25-TM run also failed in all five seeds; full results can be found in Supplementary Table~\ref{tab:sm2-autoencoder_exp_data_scaling_study}.}
\label{tab:autoencoder_exp_data_scaling_study}
\end{center}
\end{table}

\textbf{Conclusion:} a linear basis already compresses these TMs to 1\% at 62$\times$ given enough real TMs, and a from-scratch autoencoder offers nothing over it.

\subsection{1\,m vs 5\,m simulation TM comparison}
\label{1m-vs-5m-simulation-TM-comparison}

The autoencoder experiments in this section use the 124$\times$ model (latent $4 \times 13 \times 13$), the compression ratio used when this part of the study was run, with the baseline configuration (Methods), three seeds and the validation set; the fine-tuning scaling is repeated at 62$\times$ at the end of the section.

Directly simulating a 5\,m fibre is costly and mostly rejected: in 200 attempts only one TM (0.5\%) passed the condition-number filter, against 10 of 51 attempts (20\%) for 1\,m, so a kept 5\,m TM took approximately 6.9\,h against 2.2\,min for a 1\,m TM (Table~\ref{tab:1m_vs_5m_simulator_performance}). A composite 5\,m TM, the product of five 1\,m TMs, costs about 11\,min at most, and less in practice because 1\,m TMs are reused across composites.

\begin{table}[htbp]
\begin{center}
\begin{tabular}{lcc}
 \toprule
 & 1\,m & 5\,m \\
 \midrule
 Attempts & 51 & 200 \\
 Kept TMs (cond. $<$ 250) & 10 (20\%) & 1 (0.5\%) \\
 Time per attempt & $\sim$26\,s & $\sim$124\,s \\
 Total time & 0.37\,h & 6.89\,h \\
 Median condition number of attempts & $1.2\times10^{14}$ & $5.6\times10^{34}$ \\
 SVD failures (NaN/Inf) & 0 & 3 \\
 \textbf{Time per kept TM} & \textbf{$\sim$2.2\,min} & \textbf{$\sim$6.9\,h} \\
 \bottomrule
\end{tabular}
\caption{Comparing simulator performance when generating TMs for 1m and 5m fibres.}
\label{tab:1m_vs_5m_simulator_performance}
\end{center}
\end{table}

Composite TMs resemble experimental ones only loosely (Supplementary Figure~\ref{fig:SingularValues_Composite5m_vs_Exp}). The mean composite spectrum drops to about 0.3 after the first singular value and then decays slowly, with a spread between TMs of more than an order of magnitude, whereas experimental spectra fall smoothly from 1 to about 0.07 with a narrow spread.

\begin{figure}[htbp]
    \centering
    \includegraphics[width=1\textwidth]{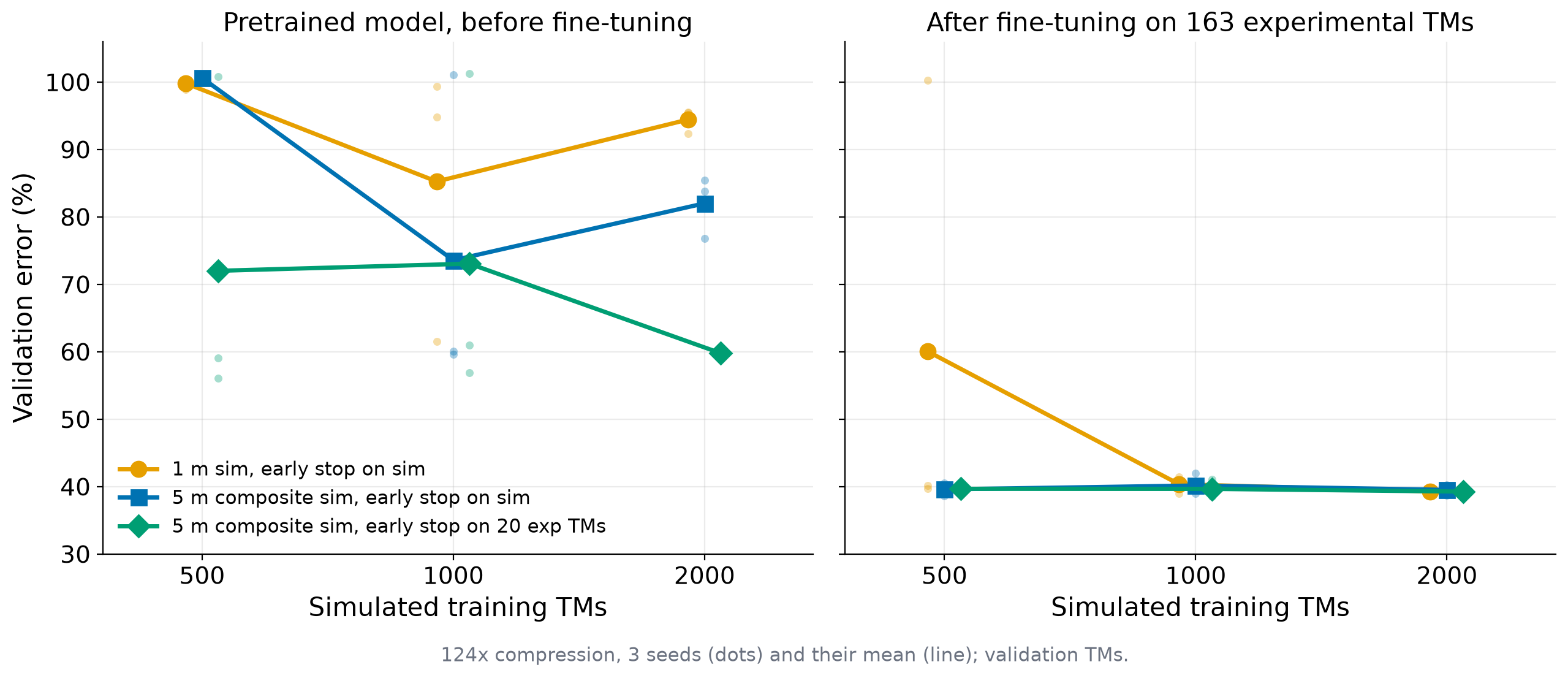}
    \caption{Effect of the simulated pretraining data on validation error at 124$\times$ compression. Left: error of the pretrained model before fine-tuning. Right: error after fine-tuning on all 163 experimental training TMs. Lines show the mean and dots the individual seeds (3 seeds), against the number of simulated training TMs. Series: 1\,m simulated TMs with early stopping on simulated validation TMs; 5\,m composite TMs with early stopping on simulated validation TMs; 5\,m composite TMs with early stopping on the 20 experimental validation TMs. Failed runs stay near 100\% error. Per-seed values and standard deviations are in Supplementary Tables~\ref{tab:sm3-1m_sim_data_scaling}--\ref{tab:sm5-5m_sim_data_scaling}.}
    \label{fig:1m_vs_composite5m_simES_and_expES}
\end{figure}

At 124$\times$ and before fine-tuning, composites give a modestly better pretrained model than 1\,m TMs: with 2,000 simulated TMs and early stopping on simulated validation TMs the error is 82.0\% for composites against 94.4\% for 1\,m, and early stopping on the 20 experimental validation TMs lowers it to $59.8 \pm 0.5$\% (Figure~\ref{fig:1m_vs_composite5m_simES_and_expES}, left). After fine-tuning on all 163 experimental TMs the three series are indistinguishable at 1,000 or more simulated TMs (39--40\%), and only 1\,m with 500 TMs fails in one seed (right). This answers Q2: composites give a useful pretraining signal, but at this compression and with the full experimental set the benefit over cheaper 1\,m simulations appears only before fine-tuning.

\textbf{Conclusion:} composite TMs are far cheaper than direct 5\,m simulation and give a better zero-shot model, but once 163 real TMs are available the choice of simulated pretraining data barely matters.

\subsubsection{Fine-tuning data scaling (124$\times$ and 62$\times$)}
\label{original_training_recipe_ends}

\begin{figure}[htbp]
    \centering
    \includegraphics[width=0.65\textwidth]{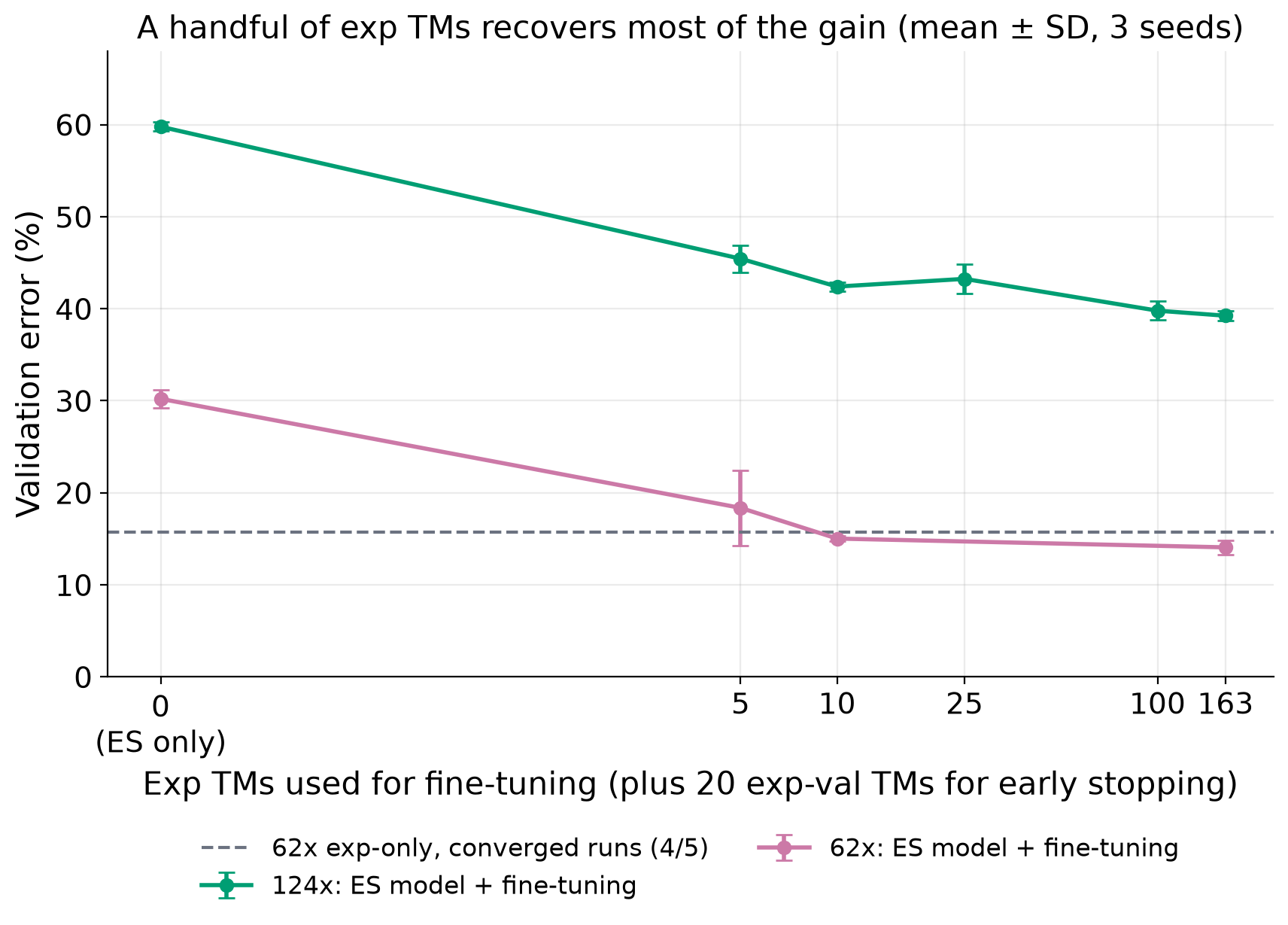}
    \caption{Validation error of the simulation-pretrained autoencoder (5\,m composite pretraining) after fine-tuning on $N$ experimental TMs, at 124$\times$ and 62$\times$ compression. $N=0$ is the pretrained model before fine-tuning. Early stopping on the 20 experimental validation TMs; baseline configuration, 3 seeds. Values are in Supplementary Tables~\ref{tab:sm6-5m_exp_data_scaling_ES_FT_124x} and~\ref{tab:sm7-5m_exp_data_scaling_ES_FT_62x}.}
    \label{fig:5m_exp_data_scaling_ES_FT_124x_62x}
\end{figure}

Here the simulation-pretrained model (5\,m composites) is fine-tuned on $N$ experimental TMs, with early stopping on the 20 experimental validation TMs (baseline recipe, 3 seeds, validation set), so the values are slightly optimistic. Most of the benefit of real TMs arrives with the first few. At 62$\times$ the pretrained-only error of $30.2 \pm 1.0$\% falls to $18.4 \pm 4.1$\% with five fine-tuning TMs and $15.0 \pm 0.3$\% with ten, against $14.1 \pm 0.8$\% with all 163; at 124$\times$ it falls from $59.8$\% to $45.4$\%, $42.4$\% and $39.2$\% (Figure~\ref{fig:5m_exp_data_scaling_ES_FT_124x_62x}; Supplementary Tables~\ref{tab:sm6-5m_exp_data_scaling_ES_FT_124x} and~\ref{tab:sm7-5m_exp_data_scaling_ES_FT_62x}). The 124$\times$ curve is not strictly monotonic ($N=25$ is slightly worse than $N=10$), but the difference is within the seed variability. This is the autoencoder half of Q4: about ten real TMs recover most of the gain from fine-tuning. It also bears on Q5: the pretrained model alone leaves a large simulation-to-experiment gap (30\% at 62$\times$), and a handful of real TMs removes roughly half of it, which a linear basis fitted on simulation cannot do (Section~\ref{Autoencoder-vs-PCA-at-62x-on-original-experimental-data}).

\textbf{Conclusion:} fine-tuning on about ten real TMs gets close to fine-tuning on all 163 at 62$\times$ ($15.0$\% against $14.1$\%).

\subsection{Autoencoder vs PCA at 62$\times$ on original experimental data}
\label{Autoencoder-vs-PCA-at-62x-on-original-experimental-data}

This section uses the final configuration, tuned using Optuna, at 62$\times$ (latent $8 \times 13 \times 13$, 1,352 reals), five seeds and the validation set. It compares the autoencoder with PCA at the same latent size, as a function of the number of real TMs available (Table~\ref{tab:ae_vs_pca_62x_exp}). The pretrained-only error ($N=0$) is $28.4 \pm 0.8$\% here, against $30.2 \pm 1.0$\% in the previous section, because the configuration and the number of seeds differ.

With few real TMs the pretrained autoencoder is the only usable model. A PCA basis fitted on 163 simulated TMs gives 90.8\% error on experimental TMs, whereas the pretrained autoencoder gives 28.4\% without seeing any real TM, and fine-tuning on five real TMs lowers this to $14.8 \pm 0.6$\%. A PCA basis fitted on the same five real TMs is limited to 724 components and gives 25.3\%. Trained from scratch on experimental TMs the autoencoder fails at $N=5$ and 10 (0/5 seeds converged) and converges in only 3/5 seeds with all 163, at $15.8 \pm 1.7$\%. This answers Q5 for this fibres: the pretrained autoencoder bridges part of the simulation-to-experiment gap, which a linear basis fitted on simulation does not (90.8\%), and the simulation-pretrained prior is what makes it work at all with very few real TMs.

The advantage ends at about ten real TMs. With $N=10$ a PCA basis reaches 1.7\% against $13.4 \pm 0.2$\% for the fine-tuned autoencoder, and with all 163 TMs 0.77\% against $11.6 \pm 0.2$\%, so PCA is about fifteen times more accurate on validation TMs at the same latent size. This answers Q3 (the nonlinear autoencoder does not compress experimental TMs better than PCA) and the crossover half of Q4 (PCA needs about ten real TMs, the pretrained autoencoder five or fewer). The autoencoder values are also optimistic, since its checkpoint was chosen on the 20 validation TMs, whereas PCA has no such selection step.

\textbf{Conclusion:} the autoencoder is not a better compressor than PCA; its value is as a simulation-pretrained prior when fewer than about ten real TMs exist.

\begin{table}[htbp]
\begin{center}
\begin{tabular}{lcccc}
 \toprule
 & \multicolumn{4}{c}{Average validation error $\pm$ SD (\%)} \\ \cmidrule(lr){2-5}
 \shortstack{Model type} & \shortstack{N=0} & \shortstack{N=5} & \shortstack{N=10} & \shortstack{N=163}\\
 \midrule
 \shortstack{AE finetuned} & $28.41 \pm 0.79$ & $14.79 \pm 0.58$ & $13.43 \pm 0.18$ & $11.59 \pm 0.23$ \\
 \lightrule
 \shortstack{AE, exp. only,\\all runs} & - & $106.73 \pm 3.95$ & $106.88 \pm 7.18$ & $49.73 \pm 46.49$ \\
 \lightrule
 \shortstack{AE, exp. only,\\converged runs} & - & none (0/5) & none (0/5) & $15.79 \pm 1.71$ (3/5) \\
 \lightrule
 PCA, sim. fit & 90.78 & - & - & - \\
 \lightrule
 PCA, exp. fit & - & 25.32$^{\dagger}$ & 1.68 & 0.77 \\
 \bottomrule
\end{tabular}
\caption{Autoencoder (AE) versus PCA at 62$\times$ compression on the original experimental TMs (validation set). $N$ is the number of experimental TMs used to fine-tune the simulation-pretrained AE ($N=0$: pretrained only, no fine-tuning), to train the AE from scratch (exp. only; ``converged'' runs have validation error below 90\%), or to fit PCA (exp. fit); PCA sim. fit uses 163 simulated TMs. AE entries are mean $\pm$ SD over five seeds and PCA entries are single deterministic fits. The AE checkpoint was chosen using the 20 validation TMs, so AE values are optimistic. Full results can be found in Supplementary Table~\ref{tab:sm8-62x_PCA_vs_AE}. $^{\dagger}$Rank-limited to 724 components.}
\label{tab:ae_vs_pca_62x_exp}
\end{center}
\end{table}

\subsection{Second fibre: zero-shot, few-shot and shared model}
\label{Collaborator-fibre:-zero-shot,-few-shot-and-shared-model}

This section asks whether the models transfer to a different, larger-core fibre (Q6). All results are at 62$\times$ with the final configuration, five seeds and the 16 second-fibre validation TMs, scored on the raw TMs and on a band-limited version (15 cycles/frame, Methods).

\subsubsection{Zero shot}
\label{Zero shot}

Without any second-fibre data the models transfer only partly (Table~\ref{tab:zero_shot_zhejiang}). On raw TMs the simulation-pretrained autoencoder gives 59.0\% error and a PCA basis fitted on simulation gives 92.4\%. Fine-tuning on original-fibre TMs does not help on the new fibre: it raises the error slightly (62--63\% for $N=5$ to 163), so adapting to one fibre does not carry over to the other. PCA fitted on 163 original TMs gives 54.6\%, slightly better than any autoencoder, and an autoencoder trained from scratch on original TMs is unreliable ($77.6 \pm 20.6$\%, with two failed seeds). Even PCA fitted on the second fibre's own 131 TMs reaches only 39.1\% at 62$\times$, so a large part of the error is a capacity limit and not only a transfer failure. On band-limited TMs the errors are lower but the picture is the same: 37.9\% for the pretrained autoencoder, 35.0\% after fine-tuning on 163 original TMs, 30.1\% for PCA fitted on original TMs, and 7.0\% for PCA fitted on the second fibre's own TMs. For Q6 the answer is therefore partial: the pretrained autoencoder transfers better than a simulation-fitted PCA, but no original-fibre model, linear or nonlinear, approaches what a model fitted on the new fibre achieves.

\textbf{Conclusion:} at a fixed 62$\times$ capacity, models built for the original fibre transfer only partially to the second fibre.

\begin{table}[htbp]
\begin{center}
\begin{tabular}{lcccc}
 \toprule
 & \multicolumn{4}{c}{Average validation error $\pm$ SD (\%)} \\ \cmidrule(lr){2-5}
 Model & $N=0$ & $N=5$ & $N=10$ & $N=163$ \\
 \midrule
 \multicolumn{5}{l}{\textit{Raw second-fibre TMs}} \\
 AE fine-tuned & $59.04 \pm 0.30$ & $62.13 \pm 0.68$ & $62.86 \pm 0.57$ & $62.62 \pm 0.35$ \\
 AE exp-only & - & - & - & $77.63 \pm 20.63$ \\
 PCA fitted on simulation & 92.43 & - & - & - \\
 PCA fitted on original TMs & - & - & - & 54.62 \\
 \midrule
 \multicolumn{5}{l}{\textit{Band-limited (15 cycles/frame)}} \\
 AE fine-tuned & $37.87 \pm 0.45$ & $35.24 \pm 0.56$ & $35.70 \pm 0.34$ & $35.01 \pm 0.59$ \\
 AE exp-only & - & - & - & $62.18 \pm 34.77$ \\
 PCA fitted on simulation & 91.15 & - & - & - \\
 PCA fitted on original TMs & - & - & - & 30.07 \\
 \bottomrule
\end{tabular}
\caption{Zero-shot transfer to the second fibre (second-fibre validation TMs, 62$\times$ compression): nothing is refitted on the second-fibre data. AE = autoencoder; $N$ is the number of original-fibre experimental TMs used to fine-tune the simulation-pretrained AE ($N=0$: pretrained only) or to fit PCA. AE entries are mean $\pm$ SD over five seeds (the exp-only mean includes two failed seeds) and PCA entries are single fits. For reference, PCA fitted on the second fibre's own 131 training TMs reaches 39.12\% (raw) and 7.02\% (band-limited), the best a 62$\times$ linear model can do on these data. Per-seed values in Supplementary Table~\ref{tab:sm9-zhejiang_zero_shot}.}
\label{tab:zero_shot_zhejiang}
\end{center}
\end{table}

\subsubsection{Few shot}
\label{Few shot}

\begin{figure}[htbp]
    \centering
    \includegraphics[width=1\textwidth]{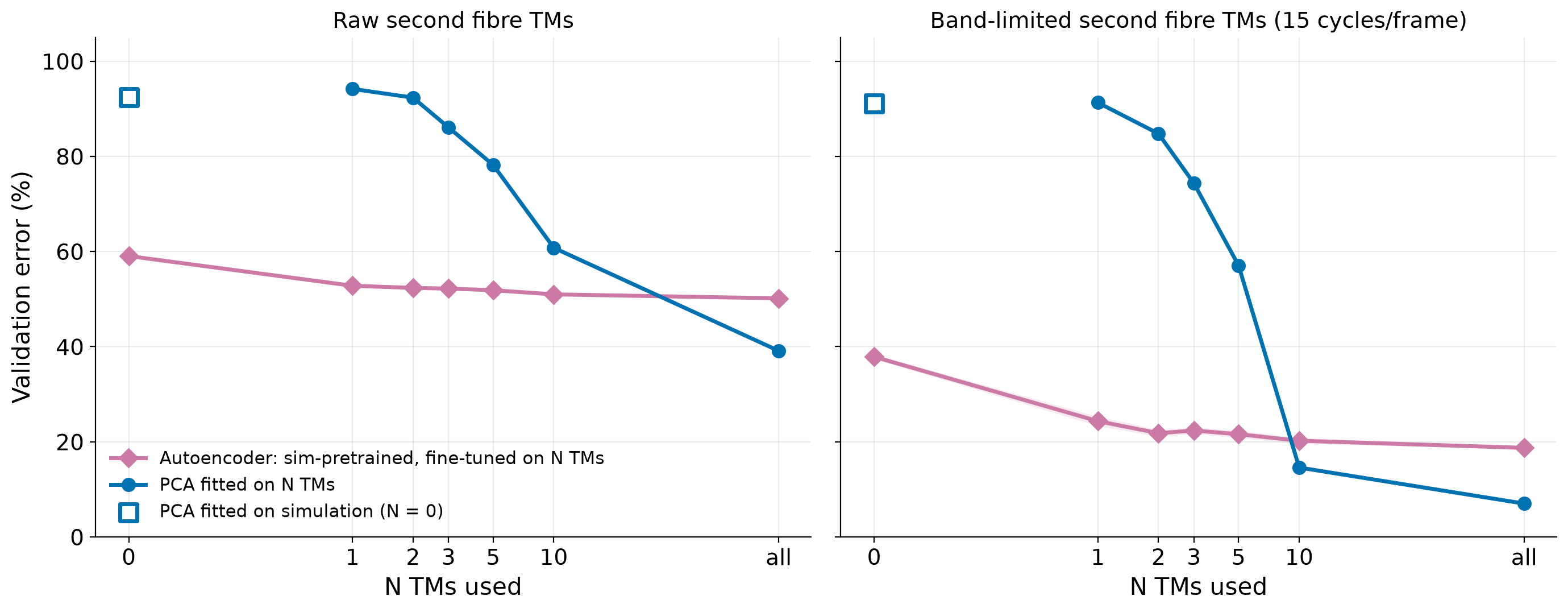}
    \caption{Few-shot calibration on the second fibre at 62$\times$ compression (second-fibre validation TMs). Left: raw TMs; right: band-limited TMs (15 cycles/frame). Autoencoder: the simulation-pretrained model fine-tuned on $N$ second-fibre TMs ($N=0$: no fine-tuning), mean $\pm$ SD over five seeds, checkpoint chosen on the second-fibre validation TMs. PCA: fitted on the same $N$ TMs (single fits), with the simulation-fitted PCA at $N=0$. ``All'' is the 131 second-fibre training TMs. Full results in Supplementary Table~\ref{tab:sm10-zhejiang_few_shot}.}
    \label{fig:few_shot_zhejiang}
\end{figure}

A single second-fibre TM already removes most of what fine-tuning can recover (Figure~\ref{fig:few_shot_zhejiang}). On raw TMs the autoencoder error falls from 59.0\% to 52.8\% with $N=1$ and 51.0\% with $N=10$, and to 50.2\% with all 131 TMs; on band-limited TMs it falls from 37.9\% to 24.3\% with $N=1$, 20.2\% with $N=10$ and 18.7\% with all 131. PCA fitted on the same few TMs is far worse (94\% and 91\% at $N=1$) and only catches up later: on raw TMs it is still behind the autoencoder at $N=10$ (60.8\% against 51.0\%) and wins only with all 131 TMs (39.1\% against 50.2\%); on band-limited TMs it overtakes the autoencoder by $N=10$ (14.6\% against 20.2\%) and reaches 7.0\% with all 131. This is the same crossover as on the original fibre (Q4), now for a second fibre (Q6): a pretrained prior helps when only a handful of target TMs exist, but a linear basis wins once about ten or more are available and the data are band-limited. The remaining autoencoder error is still large on raw TMs (about 50\%), and the extra target TMs add little beyond the first one, which again points to a capacity limit at 62$\times$.

\textbf{Conclusion:} one or two TMs from a new fibre are enough to adapt the pretrained autoencoder, but the 62$\times$ model stays above what a linear basis fitted on enough TMs of that fibre achieves.

\subsection{Shared finetuning}
\label{Shared-finetuning}

\begin{figure}[htbp]
    \centering
    \includegraphics[width=1\textwidth]{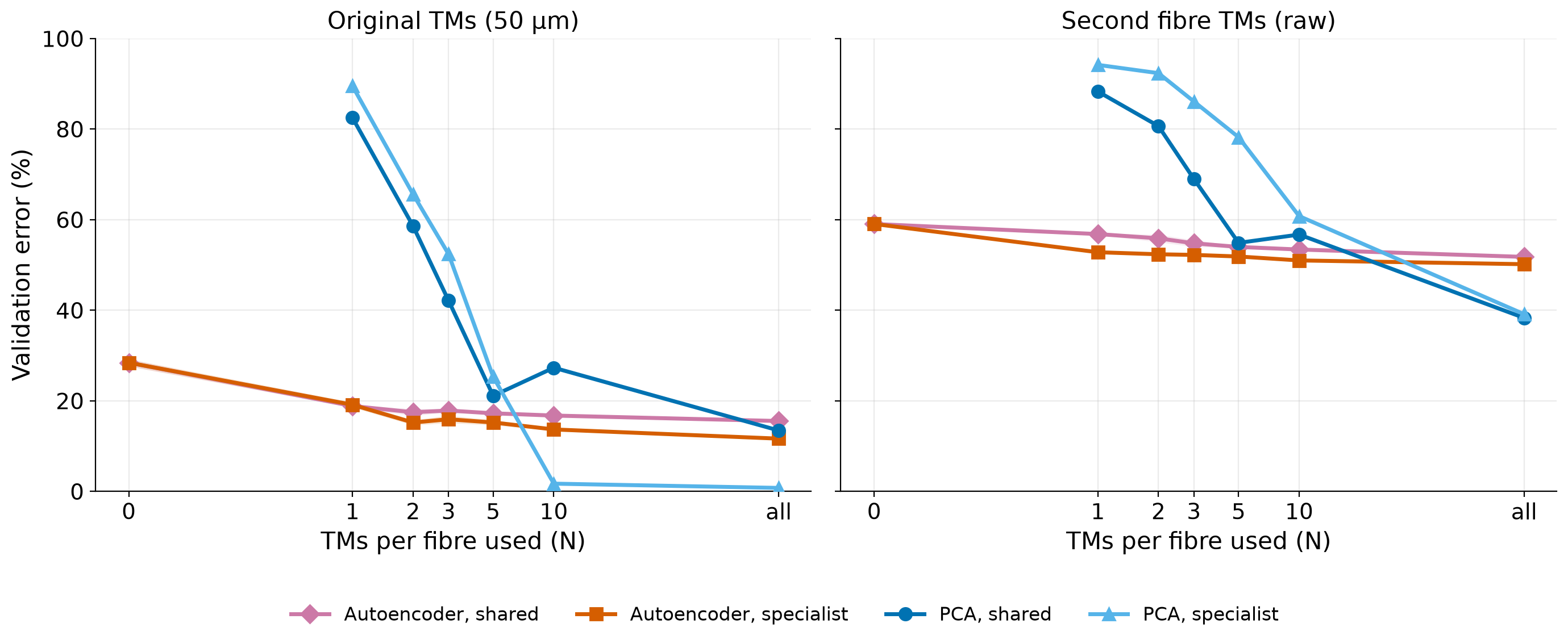}
    \caption{One shared model for both fibres versus a specialist per fibre at 62$\times$ compression (validation TMs). Shared models are fine-tuned on $N$ TMs from each fibre ($2N$ in total) and specialists on $N$ TMs of the fibre being scored; $N=0$ is the simulation-pretrained model, and ``all'' is 131 TMs per fibre (163 for the original-fibre specialist). Autoencoder: mean $\pm$ SD over five seeds, checkpoint chosen on the validation TMs; PCA: single fits. Full results in Supplementary Table~\ref{tab:sm11-shared_model}.}
    \label{fig:shared_model_test}
\end{figure}

This section asks whether a single model can serve both fibres (Q6). The simulation-pretrained autoencoder is fine-tuned on $N$ TMs from each fibre ($2N$ in total), with the checkpoint chosen on the mean of the two validation errors, and compared with a ``specialist'', a model fine-tuned on $N$ TMs of the fibre being scored only, so that it has no data from the other fibre (62$\times$, five seeds, validation set; Figure~\ref{fig:shared_model_test}). PCA fitted on the same TMs is shown for reference.

The shared autoencoder improves on the unadapted model for both fibres at every $N$, and stays close to the specialists (Figure~\ref{fig:shared_model_test}). On the second fibre it is behind the specialist at $N=1$ (56.8\% against 52.8\%) and with all TMs (51.8\% against 50.2\%). On the original fibre it is level with the specialist at $N=1$ (18.8\% against 19.1\%) and behind from $N=10$ (16.7\% against 13.7\%; 15.5\% against 11.6\% with all TMs). A single linear basis cannot do this at the same capacity: PCA fitted on both fibres reaches 13.4\% on the original fibre with all TMs, against 0.8\% for PCA fitted on that fibre alone, and its second-fibre error (38.3\%) is below the autoencoder's 51.8\%. PCA fitted on only one to three TMs of each fibre is again far worse than the autoencoder, and the two are close at $N=5$. This is the second half of Q6: one autoencoder can serve both fibres for a modest loss relative to a specialist, whereas shared PCA loses far more on the original fibre, though PCA still wins on the second fibre when many TMs are available.

\textbf{Conclusion:} a shared autoencoder serves both fibres close to a specialist for each, at the cost of an error that stays well above that of PCA when many TMs are available.

\subsection{Final test scores and robustness tests}
\label{Final-test-scores-and-robustness-tests}

This section reports the frozen test split (21 original-fibre and 17 second-fibre TMs, raw), scored once at 62$\times$. Each autoencoder entry is a single model (seed 42) and each PCA entry a single fit, so there are no error bars. It repeats the headline comparisons on data that never influenced any choice (Q3--Q6), then asks whether the remaining error matters for focusing and imaging and what survives geometric drift (Q7).

\subsubsection{Test-set errors}

The validation ordering is reproduced on the test TMs (Table~\ref{tab:final_test_results}). On the original fibre the pretrained autoencoder gives 29.2\%, two experimental TMs lower it to 14.6\% and all 163 to 11.9\%, against 90.8\% for PCA fitted on simulation, 66.7\% for PCA fitted on the same two TMs and 1.1\% for PCA fitted on all 163. The autoencoder's pretrained and 163-TM errors agree with validation (29.2\% against 28.4\%, 11.9\% against 11.6\%), so choosing its checkpoints on the 20 validation TMs did not noticeably inflate the results. These numbers confirm the answers to Q3 (PCA is more accurate at the same latent size), Q4 (the autoencoder wins with approximately five or fewer real TMs, PCA from approximately ten) and Q5 (the pretrained autoencoder closes much of the simulation-to-experiment gap that a simulation-fitted PCA cannot).

Across both fibres the shared models fall between the specialists. Fine-tuning on a fibre’s own TMs gives the lowest error (14.6\% on the original fibre, 56.7\% on the second fibre with two TMs); a shared model trained on one TM from each fibre is slightly worse (18.4\% and 61.1\%); and a specialist fine-tuned on the other fibre’s TMs is no better than the unadapted model (29.2\% against 29.2\% on the original fibre) and worse on the second fibre (67.1\% against 63.1\%). The shared model therefore avoids the failure of a mismatched specialist while being slightly worse than a matched one. With 131 TMs from each fibre, the shared autoencoder reaches 56.4\% on the second fibre, matching the two-TM specialist, whereas on the original fibre it stays above the autoencoder fine-tuned on all 163 original TMs (17.1\% against 11.9\%). As on validation, PCA is better once many TMs are available (44.3\% from the second fibre's 131 TMs, against 56.4\% for the shared autoencoder) and far worse from two (93.0\%), which answers the remainder of Q6 (Table~\ref{tab:final_test_results}).

\textbf{Conclusion:} the test split confirms the validation findings: the autoencoder helps only when almost no real TMs exist, and a shared model is slightly worse than a specialist.

\begin{table}[htbp]
\begin{center}
\begin{tabular}{lcccc}
 \toprule
 & \multicolumn{2}{c}{Autoencoder} & \multicolumn{2}{c}{PCA} \\ \cmidrule(lr){2-3}\cmidrule(lr){4-5}
 Training TMs & \shortstack{Original\\(21 TMs)} & \shortstack{Second fibre\\(17 TMs)} & \shortstack{Original\\(21 TMs)} & \shortstack{Second fibre\\(17 TMs)} \\
 \midrule
 None (simulation only)      & 29.20 & 63.05 & 90.79 & 92.94 \\
 2 original                  & 14.55 & 67.15 & 66.67 & 89.57 \\
 2 second-fibre              & 29.21 & 56.71 & 92.03 & 93.02 \\
 Shared, 1+1                 & 18.38 & 61.07 & 81.45 & 89.33 \\
 163 original                & 11.94 & 67.10 & 1.08  & 59.06 \\
 131 second-fibre            & -     & -     & 37.60 & 44.29 \\
 Shared, 131+131             & 17.09 & 56.39 & 14.31 & 43.70 \\
 \bottomrule
\end{tabular}
\caption{Final test results: error (\%) on the held-out test TMs of each fibre at 62$\times$ compression, evaluated once after all models were frozen. Autoencoder entries are single models (seed 42) and PCA entries are single deterministic fits; no model was chosen using these numbers. ``Specialist'' models use TMs of one fibre only, ``shared'' models use the stated number from each fibre, and the autoencoder checkpoints were selected on validation TMs.}
\label{tab:final_test_results}
\end{center}
\end{table}

\subsubsection{Robustness to drift}

\begin{figure}[htbp]
    \centering
    \includegraphics[width=1\textwidth]{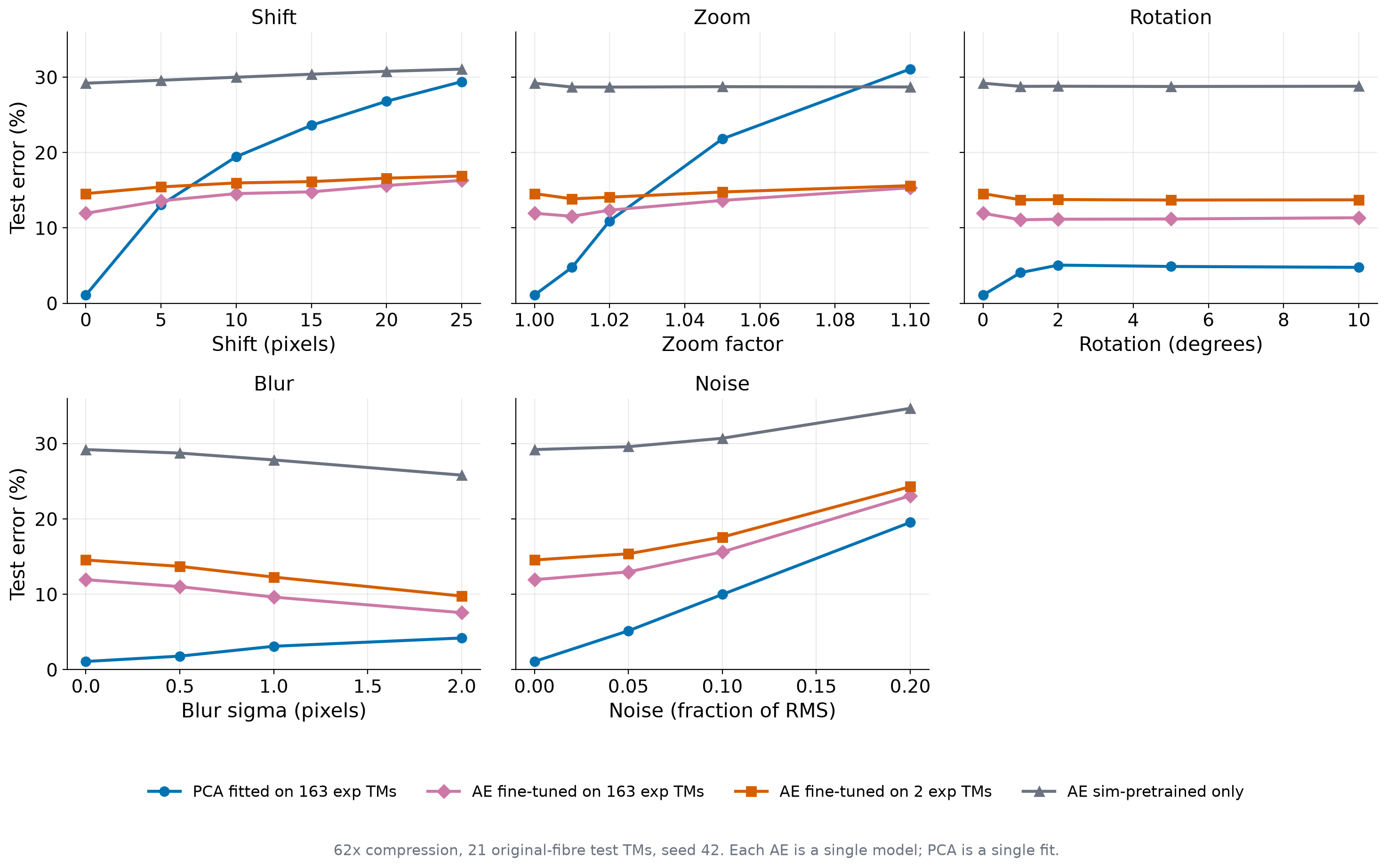}
    \caption{Test error under perturbation of the TMs before reconstruction, at 62$\times$ compression (21 original-fibre test TMs, seed 42). Shift is in pixels of the $205 \times 205$ output grid, zoom is a scale factor, rotation is in degrees, blur is the Gaussian $\sigma$ in pixels and noise is a fraction of the RMS. Each autoencoder is a single model and PCA is a single fit. Error is measured against the perturbed TM.}
    \label{fig:model_robustness_graphs}
\end{figure}

PCA is the most accurate model on unperturbed TMs, but it degrades fastest under geometric drift (Figure~\ref{fig:model_robustness_graphs}; 21 original-fibre test TMs). The PCA basis fitted on 163 TMs rises from 1.1\% to 13.1\% with a 5-pixel shift, 19.4\% with 10 pixels and 29.4\% with 25 pixels, and from 1.1\% to 10.9\% with a 2\% zoom and 21.8\% with a 5\% zoom. The autoencoder fine-tuned on 163 TMs moves only from 11.9\% to 14.6\% (10 pixels), 16.3\% (25 pixels) and 13.6\% (5\% zoom), and the two-TM specialist behaves similarly (14.6\% rising to 16.9\% at 25 pixels). PCA therefore stays ahead only for shifts up to 5 pixels and zooms up to 2\%. Over the ranges tested, PCA stays ahead of the autoencoder for rotation, blur and noise. For Q7, a linear basis fitted on aligned TMs is the better choice only if the geometry is stable, and the autoencoder is the more tolerant model; a linear basis combined with registration was not tested.

\textbf{Conclusion:} a shift of 5 pixels or a 2\% zoom is enough to erase PCA's accuracy advantage, whereas the autoencoder degrades gently.

\subsubsection{Focusing and imaging}

The reconstruction error also overstates the practical loss for the autoencoder (Figures~\ref{fig:unscramble_F} and~\ref{fig:unscramble_ring}; the target of spots is in Supplementary Figure~\ref{fig:sm-unscramble_spots}). On the original fibre the two-TM specialist reaches 98.0\% of the ideal focus enhancement (shared model 96.3\%, pretrained model 93.4\%). PCA fitted on the same two TMs reaches 53.6\% of the ideal focus, and PCA fitted on simulation 21.1\%. On the second fibre the two-TM specialist reaches 68.6\% of the ideal focus (shared 65.1\%, pretrained 63.6\%), while PCA fitted on two second-fibre TMs reaches 16.1\%. For Q7, a TM error of 12--15\% on the original fibre costs only 1--2\% of the focus, and even the 57\% error on the second fibre still reaches on average 69\% of the ideal focus, whereas a linear basis fitted on the same small budget does not.

\textbf{Conclusion:} the autoencoder's remaining reconstruction error costs little in focusing and imaging, and a linear basis with the same number of real TMs performs far worse.

\begin{figure}[htbp]
    \centering
    \includegraphics[width=1\textwidth]{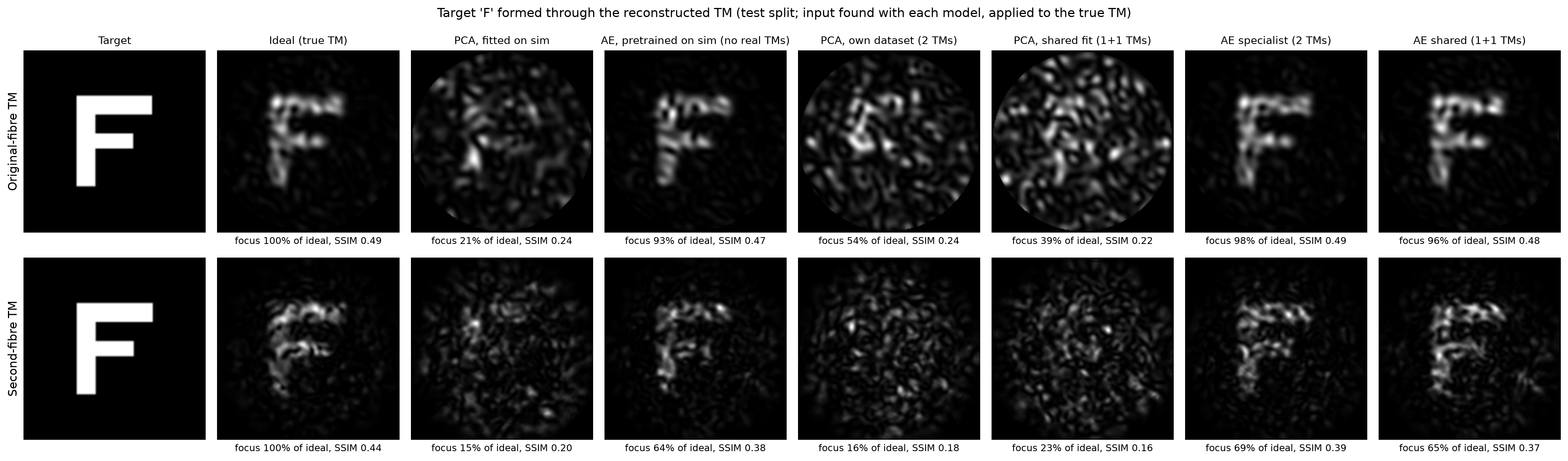}
    \caption{Focusing and imaging through reconstructed TMs for the target ``F'' (frozen test split, 62$\times$ compression). Rows: the first test TM of the original fibre and of the second fibre, the same for every model. Columns: target; ideal (the true TM, the best possible with 145 inputs); PCA fitted on 163 simulated TMs; autoencoder pretrained on simulation only; PCA fitted on two TMs of the fibre shown; PCA fitted on one TM from each fibre; autoencoder specialist (two TMs of the fibre shown); shared autoencoder (one TM from each fibre). The input is found from the reconstructed TM and applied to the true TM. Under each panel: mean focus enhancement as a percentage of the ideal and mean SSIM over all test TMs of that fibre (21 original, 17 second fibre).}
    \label{fig:unscramble_F}
\end{figure}

\begin{figure}[htbp]
    \centering
    \includegraphics[width=1\textwidth]{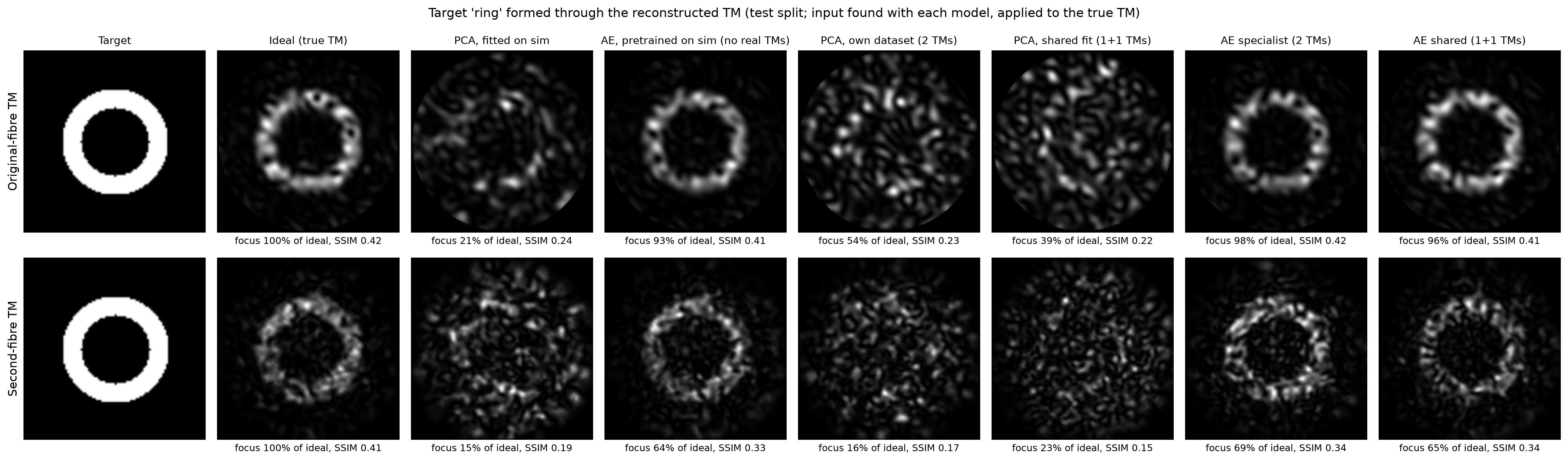}
    \caption{As Figure~\ref{fig:unscramble_F}, for the target of a ring.}
    \label{fig:unscramble_ring}
\end{figure}

\section{Discussion}
%Include a Discussion that summarizes (but does not merely repeat) your conclusions and elaborates on their implications. There should be a paragraph outlining the limitations of your results and interpretation, as well as a discussion of the steps that need to be taken for the findings to be applied. Please avoid claims of priority.

The autoencoder is not a better compressor than PCA. At 62$\times$ a linear basis fitted on the experimental TMs reaches 0.8\% validation error with 163 TMs and 1.7\% with ten, against 11.6\% for the fine-tuned autoencoder at the same latent size, and the gap persists on the frozen test TMs (1.1\% against 11.9\%) (Q1, Q3). Its value is as a prior when few or no real TMs exist. Pretrained only on simulated composite TMs produces 29\% error on experimental test TMs, where a PCA basis fitted on the same simulation gives 91\%, and 15\% after fine-tuning on two real TMs. PCA overtakes it at about ten (Q4, Q5). Pretraining also made training dependable: trained from scratch, only some seeds converged (3 of 5 with all 163 TMs in the final configuration, none with ten), whereas every seed converged after fine-tuning on 163 real TMs once 1,000 or more simulated TMs were used for pretraining (at 124$\times$ compression rate). Composite 5\,m TMs were far cheaper than direct 5\,m simulation and gave a better zero-shot model than 1\,m TMs, but once many real TMs were available the choice of simulated data barely mattered (Q2). The prior was useful even though the simulated wavelength (632.8\,nm) differed from the experimental one (660\,nm).

These results suggest when each approach is appropriate. If many TMs of a fibre can be measured, a PCA basis is simpler, faster and more accurate. If only one or two fully measured TMs can be afforded, a simulation-pretrained autoencoder adapted to them reaches 98\% of the ideal focus on the original fibre, even though its reconstruction error is 15\% (Q7). A single model can also serve two fibres at a small cost relative to a specialist model, whereas a shared linear basis loses more (Q6). The autoencoder also tolerates the shifts and zooms that quickly degrade the PCA model. A surprisingly small number of real TMs is therefore enough to adapt the model: most of the gain from fine-tuning arrives with the first five to ten. Transfer to the 105\,$\mu$m fibre was only partial. Band-limiting its TMs to the detail of the original fibre lowered the zero-shot error (38\% against 59\%), but not to the level of PCA fitted on the second fibre's own TMs (7\%), so the finer speckle of the larger core explains only part of the poor transfer.

Several limitations qualify these conclusions. All results are reconstructions of complete, measured TMs. We did not test recovering a TM from partial measurements, which would be the more demanding use. Only one source fibre (50\,$\mu$m, random bends, one setup and wavelength) and one target fibre (105\,$\mu$m, upscaled from $64 \times 64$ with the 145 brightest of 1,024 inputs) were studied, and the pretraining data covered only 50\,$\mu$m fibres. The experiment used 145 orthogonal tilted-plane-wave inputs, limited by the stability of the setup, whereas the simulations used random orthonormal inputs, which is a further difference between simulation and experiment. The autoencoder checkpoints were chosen on the 20 experimental validation TMs, which makes its validation errors optimistic, although the test errors agreed with validation. Test results are single-seed, and the robustness tests apply digital perturbations to measured TMs, without a registration step that might rescue a linear basis. The compression ratio was fixed at 62$\times$ for the transfer experiments, so poor transfer to the 105\,$\mu$m fibre partly reflects a capacity limit (even PCA fitted on that fibre reaches only 39\% raw error at this ratio), and at $N=5$ the PCA comparison is rank-limited. Only one architecture was tuned, so we cannot say that no nonlinear model could outperform PCA. Finally, we did not test pretraining sets larger than 2,000 composite TMs, so it is unknown whether more simulated data would improve performance.

Several steps would extend this work. Pretraining on a range of fibre sizes and parameters should allow one model to serve more fibres, and we expect its error on the 50\,$\mu$m fibre to rise relative to a specialist while falling on others, such as the 105\,$\mu$m fibre, but this was not tested. A larger latent size, other architectures and larger simulated sets are natural next tests. The most important extension is to recover pixel-basis TMs from fewer measurements than a full TM, which would turn the pretrained prior from a compressor into a measurement-saving tool. Compressive sampling has been demonstrated for smaller, mode-basis TMs using structural priors \cite{li2021compresstms}; to our knowledge it has not been shown for pixel-basis TMs with a learned prior.

\subsection*{Acknowledgements}

CJW: supported by the Engineering and Physical Sciences Research Council (UKRI) through a doctoral studentship. GSDG: would like to acknowledge his UKRI Future Leaders Fellowship (MR/Y034163/1 and MR/T041951/1). This research was supported by Nvidia Corporation through the Nvidia Academic Grant Program and used Nvidia RTX PRO 6000 Blackwell Max-Q Workstation Edition GPUs. We thank David B. Phillips and Robert J. Kilpatrick who jointly developed the original MATLAB fibre simulation code used in this work.

\subsection*{Author Contributions} 
%Describe contributions of each author to the paper, using the first initial and full last name. 

C.J.W.: undertook analysis, collected data, developed code, wrote manuscript. T.W.: developed and built the experimental setup to collect TMs. Z.W. \& Q.Y.: provided a second dataset of multimode fibre transmission matrices. M.P.P.: provided guidance regarding the autoencoder architecture. G.S.D.G.: supervised the project, helped formulate the ideas for the project, provided guidance, helped write manuscript.

\subsection*{Conflicts of Interest}

The author(s) declare(s) that there is no conflict of interest regarding the publication of this article.

\subsection*{Code and Data Availability}
%A data availability statement is compulsory for all research articles. This statement describes whether and how others can access the data supporting the findings of the paper, including 1) what the nature of the data is, 2) where the data can be accessed, and 3) any restrictions on data access and why.
%If data are in an archive, include the accession number or a placeholder for it. Also include any materials that must be obtained through a Material Transfer Agreements (MTA). 

Experimental TMs and RMs: Hugging Face, \url{https://doi.org/10.57967/hf/11679}. Simulated TMs: \url{https://doi.org/10.57967/hf/11677}. Code, weights and splits: GitHub, archived on Zenodo, \url{https://doi.org/10.5281/zenodo.24343908}. The second-fibre TMs are not part of these releases and can be requested from the authors who measured them (Z.W. and Q.Y.).

\printbibliography

@article{wen2023stable,
  author = {Wen, Z. and Dong, Z. and Deng, Q. and Pang, C. and Kaminski, C. F. and Xu, X. and Yan, H. and Wang, L. and Liu, S. and Tang, J. and Chen, W. and Liu, X. and Yang, Q.},
  title = {Single multimode fibre for in vivo light-field-encoded endoscopic imaging},
  journal = {Nature Photonics},
  volume = {17},
  number = {8},
  pages = {679–687},
  year = {2023},
  doi = {10.1038/s41566-023-01240-x},
  url = {https://www.nature.com/articles/s41566-023-01240-x}
}

@article{phillips2022imaging,
  author = {Phillips, D. B. and Faccio, D.},
  title = {Section 10: Imaging through multimode fibers},
  journal = {Journal of Optics},
  volume = {24},
  number = {8},
  pages = {23-25},
  year = {2022},
  doi = {10.1088/2040-8986/ac7a48},
  note = {Part of: Ilaria Cristiani et al 2022 J. Opt. 24 083001},
  url = {https://iopscience.iop.org/article/10.1088/2040-8986/ac7a48}
}

@inproceedings{wilson2026bend,
    author = {Wilson, C. J. and Zheng, Y. and Phillips, D. B. and Gordon, G. S. D.},
    title = {Using physics-informed AI to enable fibre TM recovery from minimal partial measurements enabling ultra-thin endoscopes},
    booktitle = {Proc. SPIE 13865, Computational Optical Imaging and Artificial Intelligence in Biomedical Sciences III, 138650A},
    year =  {2026},
    doi = {10.1117/12.3077568}
}

@article{popoff2010imagetransmission,
  author = {Popoff, S. and Lerosey, G. and Fink, M. and  Boccara, A. C. and Gigan, S.},
  title = {Image transmission through an opaque material},
  journal = {Nature Communications},
  volume = {1},
  number = {81},
  year = {2010},
  doi = {10.1038/ncomms1078},
  url = {https://www.nature.com/articles/ncomms1078}
}

@article{zheng2023single,
  author = {Zheng, Y. and Wright, T. and Wen, Z. and Yang, Q. and Gordon, G. S. D.},
  title = {Single-ended recovery of optical fiber transmission matrices using neural networks},
  journal = {Communications Physics},
  volume = {6},
  number = {1},
  pages = {306},
  year = {2023},
  doi = {10.1038/s42005-023-01410-x},
  url = {https://www.nature.com/articles/s42005-023-01410-x}
}

@article{gordon2024flexible,
  author = {Gordon, G. S. D.},
  title = {Flexible ultra-thin super-resolution endoscopy},
  journal = {Science China - Physics Mechanics \& Astronomy},
  volume = {67},
  number = {3},
  pages = {234232},
  year = {2024},
  doi = {10.1007/s11433-023-2237-5},
  url = {https://link.springer.com/article/10.1007/s11433-023-2237-5}
}

@article{amitonova2024MMFBrain,
  author = {Amitonova, L. V.},
  title = {Multimode fiber endoscopes for computational brain imaging},
  journal = {Neurophotonics},
  volume = {11},
  number = {1},
  pages = {S11509},
  year = {2024},
  doi = {10.1117/1.NPh.11.S1.S11509},
  note = {Accessed: May 20, 2025},
  url = {https://www.spiedigitallibrary.org/journals/neurophotonics/volume-11/issue-S1/S11509/Multimode-fiber-endoscopes-for-computational-brain-imaging/10.1117/1.NPh.11.S1.S11509.full}
}

@article{kim2015transmission,
  author = {Kim, M. and Choi, W. and Choi, Y. and Yoon, C. and Choi, W.},
  title = {Transmission matrix of a scattering medium and its applications in biophotonics},
  journal = {Optics Express},
  volume = {23},
  number = {10},
  pages = {12648--12668},
  year = {2015},
  doi = {10.1364/OE.23.012648},
  url = {https://opg.optica.org/oe/fulltext.cfm?uri=oe-23-10-12648&id=318425}
}

@article{mienye2025ae-review,
  author = {Mienye, I. D. and Swart, T. G.},
  title = {Deep Autoencoder Neural Networks: A Comprehensive Review and New Perspectives},
  journal = {Archives of Computational Methods in Engineering},
  volume = {32},
  pages = {3981–-4000},
  year = {2025},
  doi = {10.1007/s11831-025-10260-5},
}

@article{wang2022compress-ae,
  author = {Wang, N. and Liu, T. and Wang, J. and Liu, Q. and Alibhai, S. and He, X.},
  title = {Locality-based transfer learning on compression autoencoder for efficient scientific data lossy compression},
  journal = {Journal of Network and Computer Applications},
  volume = {205},
  number = {103452},
  year = {2022},
  doi = {10.1016/j.jnca.2022.103452},
}

@misc{zheng2024attentiontmlatent,
  author = {Zheng, Y. and Kilpatrick, R. J. and Phillips, D. B. and Gordon, G. S. D.},
  title = {Self-attention networks enable compact latent space modelling of dynamic optical fibre matrices},
  year = {2024},
  eprint = {2406.07775},
  archivePrefix = {arXiv},
  primaryClass = {cs.LG},
  doi = {10.48550/arXiv.2406.07775}
}

@inproceedings{zheng2025efficient-ai-tm,
    author = {Zheng, Y. and Kilpatrick, R. J. and Yang, Q. and Phillips, D. B. and Gordon, G. S. D.},
    title = {Adapting the ChatGPT architecture for efficient AI representation and modelling of optical fibre transmission matrices},
    booktitle = {25th Anniversary International Conference on Transparent Optical Networks (ICTON)},
    year =  {2025},
    doi = {10.1109/ICTON67126.2025.11125157}
}

@article{gordon2019metasurface,
  author = {Gordon, G. S. D. and Gataric, M. and Ramos, A. G. C. P. and Mouthaan, R. and Williams, C. and Yoon, J. and Wilkinson, T. D. and Bohndiek, S. E.},
  title = {Characterizing Optical Fiber Transmission Matrices Using Metasurface Reflector Stacks for Lensless Imaging without Distal Access},
  journal = {Physical Review X},
  volume = {9},
  number = {4},
  pages = {041050},
  year = {2019},
  doi = {10.1103/PhysRevX.9.041050},
}

@article{argyros2008tmm,
  author = {Argyros, A. and Lwin, R. and  Large, M. C. J.},
  journal = {Optics Express},
  title = {Bend loss in highly multimode fibres},
  volume = {16},
  number = {23},
  pages = {18590-18598},
  year = {2008},
  note = {Accessed: Jun. 16, 2025},
  url = {https://opg.optica.org/oe/fulltext.cfm?uri=oe-16-23-18590&id=173060},
  doi = {10.1364/OE.16.018590}
}

@software{popoff2024pymmf,
  author       = {Popoff, Sébastien M.},
  title        = {wavefrontshaping/pyMMF: 0.7},
  year         = {2024},
  publisher    = {Zenodo},
  doi          = {10.5281/zenodo.13763018},
  version      = {0.7},
}

@article{veettikazhy2021bpm,
  author = {Veettikazhy, M. and Hansen, A. K. and Marti, D. and Jensen, S. M. and Borre, A. L. and Andresen, E. R. and Dholakia, K. and Andersen, P. E.},
  title = {BPM-Matlab: an open-source optical propagation simulation tool in MATLAB},
  journal = {Optics Express},
  volume = {29},
  number = {8},
  pages = {11819--11832},
  year = {2021},
  doi = {10.1364/OE.420493}
}

@article{feng2026diffusion,
  author = {Feng, Z. and Fan, H. and Liu, H. and Han, T. and Wang, J. and Tan, Y. and Zhou, W. and Xu, B. and Liu, J. and Zheng, J. and Sun, J. and Xiong, D. and Yang, X.},
  title = {Physics-informed diffusion enables high-fidelity multimode fiber imaging under dynamic perturbations},
  journal = {Commun Phys},
  volume = {9},
  number = {324},
  year = {2026},
  doi = {10.1038/s42005-026-02728-y}
}

@article{ma2026rvitm-biophotonics,
  author = {Ma, H. and Gautam, R. and Andersson-Engels, S. and Sekar S. K. V.},
  title = {Real-valued intensity transmission matrix-based multimode fibre beamshaping for biophotonics applications},
  journal = {Opt. Lett.},
  volume = {51},
  pages = {4976-4979},
  year = {2026},
  doi = {10.1364/OL.603130}
}

@article{skvarenina2025bend,
  author = {Skvarenina, L. and Simpson, S. and Alizadeh, Y. and Lavery, M. P. J.},
  title = {Extraction of bend-resolved modal basis in deformed multimode fiber},
  journal = {APL Photonics},
  volume = {10},
  number = {8},
  pages = {086108},
  year = {2025},
  doi = {10.1063/5.0262067}
}

@article{hinton2006reducing,
  author = {Hinton, G. E. and Salakhutdinov, R. R.},
  title = {Reducing the Dimensionality of Data with Neural Networks},
  journal = {Science},
  volume = {313},
  pages = {504-507},
  year = {2006},
  doi = {10.1126/science.1127647}
}

@inproceedings{optuna_2019,
	title={Optuna: A Next-generation Hyperparameter Optimization Framework},
	author={Akiba, Takuya and Sano, Shotaro and Yanase, Toshihiko and Ohta, Takeru and Koyama, Masanori},
	booktitle={Proceedings of the 25th {ACM} {SIGKDD} International Conference on Knowledge Discovery and Data Mining},
	year={2019},
  doi = {10.1145/3292500.3330701}
}

@misc{wang2020linformer,
  author = {Wang, S. and Li, B. Z. and Khabsa, M. and Fang, H. and Ma, H.},
  title = {Linformer: Self-Attention with Linear Complexity},
  year = {2020},
  eprint = {2006.04768},
  archivePrefix = {arXiv},
  primaryClass = {cs.LG},
  note = {Accessed: Jun. 08, 2026},
  doi = {10.48550/arXiv.2006.04768}
}

@article{li2021compresstms,
  author = {Li, S. and Saunders, C. and Lum, D. J. and Murray-Bruce, J. and Goyal, V. K. and Čižmár, T. and Phillips, D. B.},
  title = {Compressively sampling the optical transmission matrix of a multimode fibre},
  journal = {Light: Science \& Applications},
  volume = {10},
  number = {88},
  year = {2021},
  doi = {10.1038/s41377-021-00514-9}
}

@article{gordon2019fullfield,
  author = {Gordon, G. S. D. and Joseph, J. and Sawyer, T. and Macfaden, A. J. and Williams, C. and Wilkinson, T. D. and Bohndiek, S. E.},
  title = {Full-field quantitative phase and polarisation-resolved imaging through an optical fibre bundle},
  journal = {Optics Express},
  volume = {27},
  number = {17},
  pages = {23929--23947},
  year = {2019},
  doi = {10.1364/OE.27.023929}
}

@book{goodman1996fourier,
  author    = {Goodman, Joseph W.},
  title     = {Introduction to {F}ourier Optics},
  edition   = {2},
  publisher = {McGraw-Hill},
  year      = {1996}
}

@article{cizmar2011shapinglight,
  author = {Čižmár, T. and Dholakia, K.},
  title = {Shaping the light transmission through a multimode optical fibre: complex transformation analysis and applications in biophotonics},
  journal = {Optics Express},
  volume = {19},
  number = {20},
  pages = {18871--18884},
  year = {2011},
  doi = {10.1364/OE.19.018871}
}

@misc{kilpatrick2026adaptive,
  author       = {Kilpatrick, R. J. and Milne, A. and Harrington, K. and Wright, T. A. and Rocha, J. C. A. and Beno{\^\i}t, A. and Parker, H. and Mididoddi, C. K. and Stone, J. M. and Birks, T. A. and Thomson, R. R. and Phillips, D. B.},
  title        = {Adaptive wavefront shaping through flexible multicore photonic lantern fibres},
  year         = {2026},
  eprint       = {2611.10290},
  archivePrefix = {arXiv},
  primaryClass = {physics.optics},
  doi = {10.48550/arXiv.2611.10290}
}

@article{xiong2020dlmmf,
  author = {Xiong, W. and Redding, B. and Gertler, S. and Bromberg, Y. and Tagare, H. D. and Cao, H.},
  title = {Deep learning of ultrafast pulses with a multimode fiber},
  journal = {APL Photonics},
  volume = {5},
  number = {9},
  pages = {096106},
  year = {2020},
  doi = {10.1063/5.0007037}
}

\clearpage
\setcounter{table}{0}\setcounter{figure}{0}
\renewcommand{\thetable}{S\arabic{table}}
\renewcommand{\thefigure}{S\arabic{figure}}
\renewcommand{\theHtable}{S\arabic{table}}
\renewcommand{\theHfigure}{S\arabic{figure}}
% Compact layout for the supplement only (everything below this point)
\setstretch{1.0}
\renewcommand{\arraystretch}{0.9}
\AtBeginEnvironment{tabular}{\footnotesize}
\captionsetup{font=footnotesize}
\setlength{\floatsep}{8pt plus 2pt minus 2pt}
\setlength{\textfloatsep}{10pt plus 2pt minus 2pt}
\setlength{\intextsep}{8pt plus 2pt minus 2pt}
\renewcommand{\topfraction}{0.9}
\renewcommand{\bottomfraction}{0.9}
\renewcommand{\textfraction}{0.1}
\renewcommand{\floatpagefraction}{0.8}
% Top-align figures on float-only pages (LaTeX centres them vertically by default)
\makeatletter
\setlength{\@fptop}{0pt}
\setlength{\@fpsep}{8pt plus 1fil}
\setlength{\@fpbot}{0pt plus 1fil}
\makeatother
\section*{Supplementary Material}

Unless stated otherwise, tables report validation errors (\%) for the original experimental data. Mean $\pm$ SD are over seeds; failed runs stay near 100\%.

\begin{table}[htbp]
\begin{center}
\small
\setlength{\tabcolsep}{4pt}
\begin{tabular}{llcccc}
 \toprule
 Model & Frozen during fine-tuning & PT sim & PT exp & FT sim & FT exp \\
 \midrule
 Full attention & enc, to\_latent, channel\_attn & 37.31 & 68.31 & 382.49 & 52.98 \\
 & enc, to\_latent & & & 312.78 & 52.53 \\
 & none & & & 519.30 & 44.56 \\
 \midrule
 Coarse (Linformer) attention & enc, to\_latent, channel\_attn & 35.76 & 70.81 & 792.54 & 52.85 \\
 & enc, to\_latent & & & 1223.04 & 52.26 \\
 & none & & & 412.66 & 45.26 \\
 \midrule
 \shortstack[l]{1D convolutions\\+ coarse attention} & enc, to\_latent, channel\_attn & 99.07 & 100 & 99.04 & 100 \\
 & enc, to\_latent & & & 99.04 & 100 \\
 & none & & & 99.04 & 100 \\
 \bottomrule
\end{tabular}
\caption{Preliminary comparison of attention variants on smaller $256 \times 256$ TMs, run before the experimental data were collected; each model was tuned with 50 Optuna trials. PT = pretrained on simulated TMs; FT = after fine-tuning on experimental TMs; ``sim'' and ``exp'' give the phase-aligned error (\%) on simulated and experimental validation TMs. FT exp is the metric of interest; FT sim is not, since the model moves away from the simulated data when fine-tuned. Freezing layers affects only the FT columns. Single runs.}
\label{tab:sm-attention_variants}
\end{center}
\end{table}

\begin{table}[htbp]
\begin{center}
\small
\setlength{\tabcolsep}{6pt}
\begin{tabular}{lccc}
 \toprule
 & & \multicolumn{2}{c}{Validation error (\%)} \\ \cmidrule(lr){3-4}
 Model (latent shape per speckle) & Latent reals & Pretrained & Fine-tuned \\
 \midrule
 Two-channel, $16 \times 13 \times 13$ (real) & 2{,}704 & 258.13 & 7.47 \\
 Complex-valued, $16 \times 13 \times 13$ (complex) & 5{,}408 & 34.70 & 13.13 \\
 Complex-valued, $8 \times 13 \times 13$ (complex) & 2{,}704 & 111.78 & 101.20 \\
 \bottomrule
\end{tabular}
\caption{Preliminary comparison of a complex-valued and a two-channel autoencoder with the same structure (channel attention, 2D convolutional encoder, residual attention) on the pixel-basis TMs, scored on the experimental validation TMs after pretraining on simulated TMs and after fine-tuning on experimental TMs. A complex latent value counts as two real numbers, so the two-channel model with 16 real channels (about 31$\times$ compression) is matched in latent size by the complex model with 8 complex channels, and the complex model with 16 channels has twice the latent size (about 15.5$\times$). The complex model uses complex weights with a modReLU activation. The relative error is not bounded by 100\%, so values above 100\% are possible. Single runs under an earlier protocol.}
\label{tab:sm-complex_vs_2channel}
\end{center}
\end{table}
\begin{figure}[htbp]
    \centering
    \includegraphics[width=0.8\textwidth]{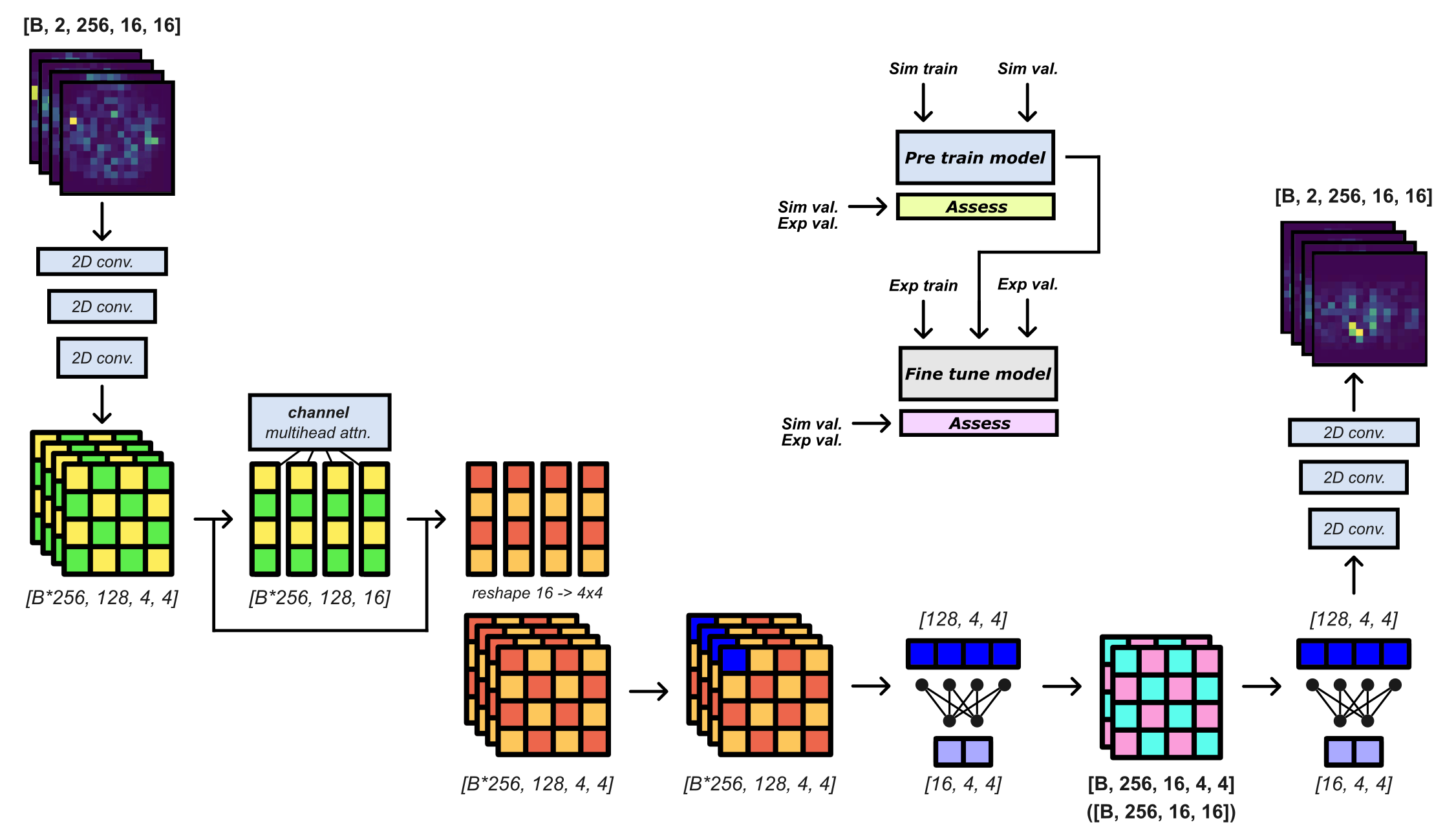}
    \caption{Full channel-attention autoencoder from the preliminary tests on $256 \times 256$ TMs: each speckle is passed through a 2D convolutional encoder, channel self-attention with a residual connection, a linear layer to the latent code, an upscaling linear layer, plus a 2D convolutional decoder. The inset shows the pretraining and fine-tuning procedure.}
    \label{fig:sm-attn_full}
\end{figure}

\begin{figure}[htbp]
    \centering
    \includegraphics[width=0.8\textwidth]{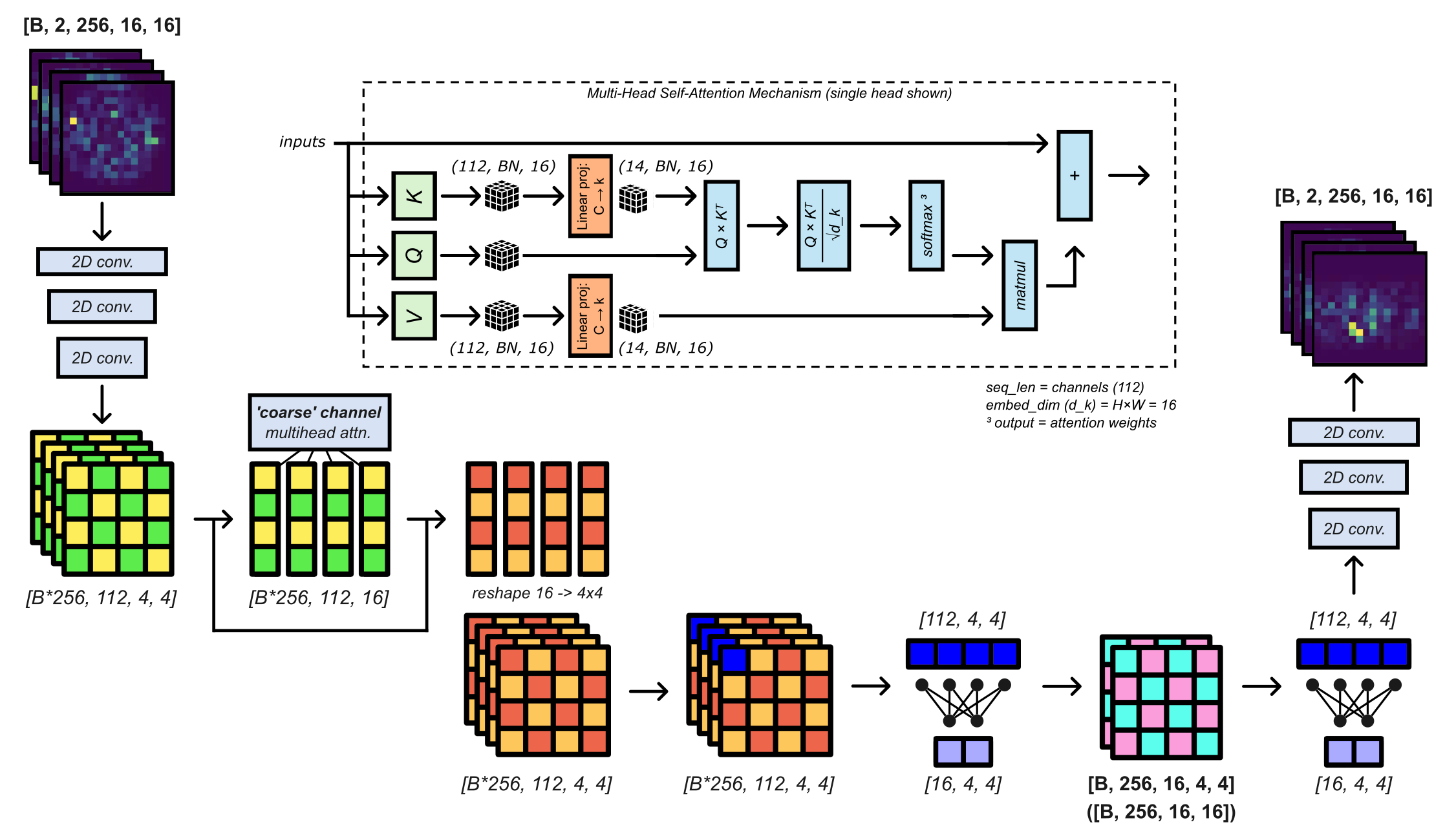}
    \caption{Coarse (Linformer-style) channel attention: as in the previous figure, but the keys and values are linearly projected from 112 channels to 14 before the attention product, which reduces the attention matrix from $112 \times 112$ to $112 \times 14$. The model was tuned separately, so its channel count differs from the full-attention model.}
    \label{fig:sm-attn_coarse}
\end{figure}

\begin{figure}[htbp]
    \centering
    \includegraphics[width=0.8\textwidth]{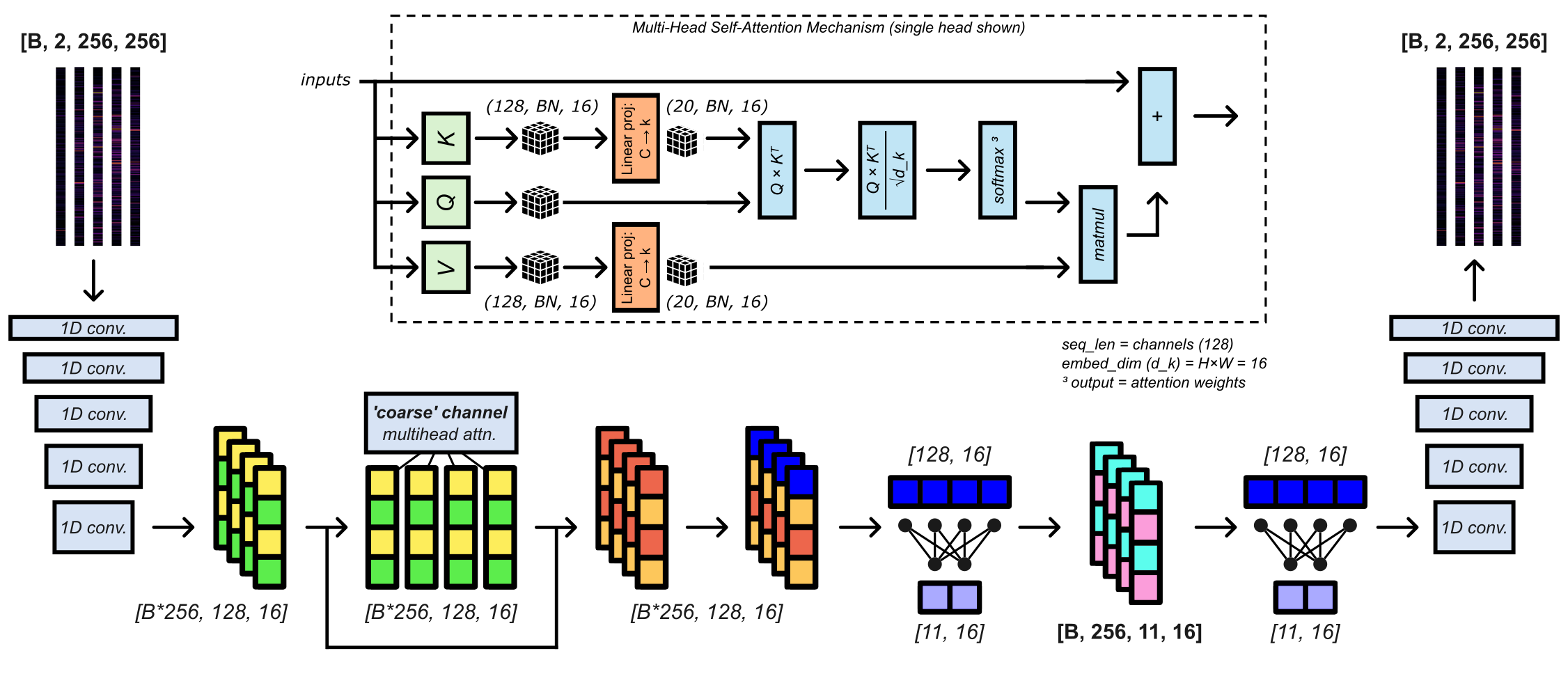}
    \caption{One-dimensional convolutional variant with coarse attention: each speckle is flattened to a 256-element vector and encoded with 1D convolutions. This model failed to train in all configurations (Supplementary Table S1).}
    \label{fig:sm-attn_1dconv}
\end{figure}

\begin{figure}[htbp]
    \centering
    \includegraphics[width=0.65\textwidth]{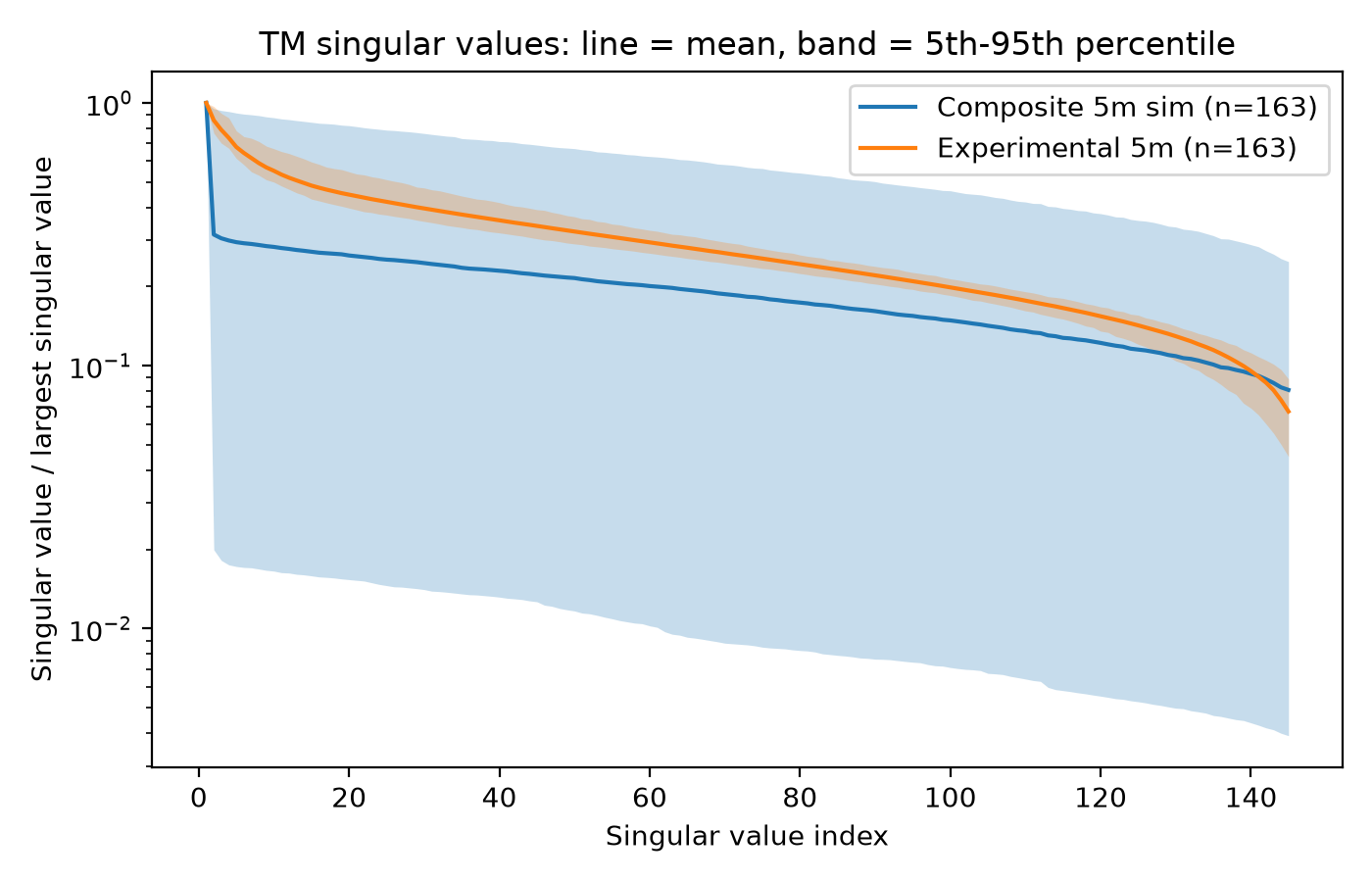}
    \caption{Singular values of composite 5\,m TMs compared with experimental TMs.}
    \label{fig:SingularValues_Composite5m_vs_Exp}
\end{figure}

\begin{table}[htbp]
\begin{center}
\small
\setlength{\tabcolsep}{4pt}
\begin{tabular}{ccccccccc}
 \toprule
 & & \multicolumn{7}{c}{Validation error (\%)} \\ \cmidrule(lr){3-9}
 \shortstack{Compression\\rate} & \shortstack{Seeds\\converged} & \shortstack{Converged\\avg. $\pm$ SD} & \shortstack{All-seed\\avg. $\pm$ SD} & \shortstack{seed\\42} & \shortstack{seed\\43} & \shortstack{seed\\44} & \shortstack{seed\\45} & \shortstack{seed\\46} \\
 \midrule
 16$\times$ & 2/5 & 12.03 $\pm$ 2.64 & 64.97 $\pm$ 48.35 & 100.18 & 13.89 & 100.15 & 100.48 & 10.16 \\
 31$\times$ & 2/5 & 11.62 $\pm$ 1.32 & 64.99 $\pm$ 48.73 & 12.55 & 10.68 & 100.50 & 100.52 & 100.72  \\
 62$\times$ & 4/5 & 15.70 $\pm$ 2.00 & 32.64 $\pm$ 37.93 & 100.43 & 14.37 & 18.67 & 14.77 & 14.97\\
 83$\times$ & 1/5 & 27.22 & 85.58 $\pm$ 32.62  & 100.08 & 100.02 & 27.22 & 100.15 & 100.42\\
 124$\times$ & 0/5 & n/a & 100.27 $\pm$ 0.07 & 100.26 & 100.24 & 100.37 & 100.19 & 100.31\\
 \bottomrule
\end{tabular}
\caption{Autoencoder experimental only compression study: per seed results.}
\label{tab:sm1-autoencoder_exp_compression_study}
\end{center}
\end{table}

\begin{table}[htbp]
\begin{center}
\small
\setlength{\tabcolsep}{4pt}
\begin{tabular}{ccccccccc}
 \toprule
 & & \multicolumn{7}{c}{Validation error (\%)} \\ \cmidrule(lr){3-9}
 \shortstack{N. training\\TMs} & \shortstack{Seeds\\converged} & \shortstack{Converged\\avg. $\pm$ SD} & \shortstack{All-seed\\avg. $\pm$ SD} & \shortstack{seed\\42} & \shortstack{seed\\43} & \shortstack{seed\\44} & \shortstack{seed\\45} & \shortstack{seed\\46} \\
 \midrule
 25 &  0/5 & n/a & 100.78 $\pm$ 0.38 & 100.74 & 101.24 & 100.98 & 100.76 & 100.20 \\
 100 & 0/5 & n/a & 100.89 $\pm$ 1.50 & 103.57 & 100.36 & 100.15 & 100.15 & 100.22\\
 125 & 0/5 & n/a & 100.57 $\pm$ 0.55 & 100.51 & 100.20 & 101.53 & 100.26 & 100.35\\
 150 & 2/5 & 13.59 $\pm$ 0.76 & 65.65 $\pm$ 47.53 & 100.77 & 14.13 & 100.13 & 13.05 & 100.18\\
 163 & 4/5 & 15.70 $\pm$ 2.00 & 32.64 $\pm$ 37.93 & 100.43 & 14.37 & 18.67 & 14.77 & 14.97\\
 \bottomrule
\end{tabular}
\caption{Autoencoder experimental-only data scaling study, with fixed compression rate of 62$\times$: per seed results.}
\label{tab:sm2-autoencoder_exp_data_scaling_study}
\end{center}
\end{table}

\begin{table}[htbp]
\begin{center}
\begin{tabular}{cccc}
 \toprule
 & & \multicolumn{2}{c}{Validation error (\%)} \\ \cmidrule(lr){3-4}
 N. sim. training TMs & Seed & Zero-shot $\pm$ SD & Fine-tuned $\pm$ SD \\
 \midrule
 500 & Avg. & $99.82 \pm 0.83$ & $60.03 \pm 34.82$ \\
 & 42 & 100.30 & 100.24 \\
 & 43 & 98.86 & 39.73 \\
 & 44 & 100.29 & 40.12 \\
 1000 & Avg. & $85.24 \pm 20.67$ & $40.34 \pm 1.26$ \\
 & 42 & 61.52 & 41.46 \\
 & 43 & 94.82 & 38.98 \\
 & 44 & 99.38 & 40.58 \\
 2000 & Avg. & $94.42 \pm 1.76$ & $39.28 \pm 0.20$ \\
 & 42 & 95.39 & 39.20 \\
 & 43 & 95.48 & 39.13 \\
 & 44 & 92.38 & 39.50 \\
 \bottomrule
\end{tabular}
\caption{Data scaling law for 1\,m simulated fibre TMs using autoencoder with 124$\times$ compression: per seed results.}
\label{tab:sm3-1m_sim_data_scaling}
\end{center}
\end{table}

\begin{table}[htbp]
\begin{center}
\begin{tabular}{cccc}
 \toprule
 & & \multicolumn{2}{c}{Validation error (\%)} \\ \cmidrule(lr){3-4}
 N. sim. training TMs & Seed & Zero-shot $\pm$ SD & Fine-tuned $\pm$ SD \\
 \midrule
 500 & Avg. & $100.60 \pm 0.33$ & $39.58 \pm 0.93$ \\
 & 42 & 100.85 & 40.48 \\
 & 43 & 100.72 & 38.63 \\
 & 44 & 100.22 & 39.62 \\
 1000 & Avg. & $73.56 \pm 23.82$ & $40.15 \pm 1.57$ \\
 & 42 & 59.57 & 41.93 \\
 & 43 & 101.06 & 39.00 \\
 & 44 & 60.05 & 39.51 \\
 2000 & Avg. & $82.00 \pm 4.62$ & $39.50 \pm 0.77$ \\
 & 42 & 76.75 & 40.23 \\
 & 43 & 83.82 & 38.69 \\
 & 44 & 85.43 & 39.58 \\
 \bottomrule
\end{tabular}
\caption{Data scaling law for 5\,m simulated fibre TMs using autoencoder with 124$\times$ compression: per seed results.}
\label{tab:sm4-5m_sim_data_scaling}
\end{center}
\end{table}

\begin{table}[htbp]
\begin{center}
\begin{tabular}{cccc}
 \toprule
 & & \multicolumn{2}{c}{Validation error (\%)} \\ \cmidrule(lr){3-4}
 N. sim. training TMs & Seed & Zero-shot $\pm$ SD & Fine-tuned $\pm$ SD \\
 \midrule
 500 & Avg. & $71.99 \pm 24.99$ & $39.66 \pm 0.78$ \\
 & 42 & 56.08 & 40.31 \\
 & 43 & 59.10 & 38.80 \\
 & 44 & 100.79 & 39.86 \\
 1000 & Avg. & $73.05 \pm 24.53$ & $39.65 \pm 1.30$ \\
 & 42 & 60.98 & 39.39 \\
 & 43 & 56.90 & 38.49 \\
 & 44 & 101.28& 41.06 \\
 2000 & Avg. & $59.78 \pm 0.50$ & $39.24 \pm 0.50$ \\
 & 42 & 59.46 & 39.77 \\
 & 43 & 59.52 & 38.77 \\
 & 44 & 60.36 & 39.19 \\
 \bottomrule
\end{tabular}
\caption{Data scaling law for 5\,m simulated fibre TMs using autoencoder with 124$\times$ compression. Experimental data used for early stopping: per seed results.}
\label{tab:sm5-5m_sim_data_scaling}
\end{center}
\end{table}

\begin{table}[htbp]
\begin{center}
\begin{tabular}{cccc} 
 \toprule
 & & \multicolumn{2}{c}{Validation error (\%)} \\ \cmidrule(lr){3-4}
 \shortstack{N experimental\\finetuning TMs} & Seed & \shortstack{Avg. finetuned error\\$\pm$ SD} & \shortstack{Avg. gain vs\\pretraining$\pm$ SD} \\
 \midrule
 0 & Avg. & $59.78 \pm 0.50$ & - \\
 & 42 & 59.46 & - \\
 & 43 & 59.52 & - \\
 & 44 & 60.36 & - \\
 5 & Avg. & $45.43 \pm 1.47$ & 14.35 \\
 & 42 & 45.45 & 14.01 \\
 & 43 & 43.96 & 15.56 \\
 & 44 & 46.89 & 13.47 \\
 10 & Avg. & $42.41 \pm 0.48$ & 17.37 \\
 & 42 & 42.03 & 17.43 \\
 & 43 & 42.24 & 17.28 \\
 & 44 & 42.95 & 17.41 \\
 25 & Avg. & $43.25 \pm 1.60$ & 16.53 \\
 & 42 & 45.04 & 14.42 \\
 & 43 & 42.73 & 16.79 \\
 & 44 & 41.97 & 18.39 \\
 100 & Avg. & $39.79 \pm 1.01$ & 19.99\\
 & 42 & 40.07 & 19.39 \\
 & 43 & 38.67 & 20.85 \\
 & 44 & 40.63 & 19.73 \\
 163 & Avg. & $39.24 \pm 0.50$ & 20.54\\
 & 42 & 39.77 & 19.69 \\
 & 43 & 38.77 & 20.75 \\
 & 44 & 39.19 & 21.17 \\
 \bottomrule
\end{tabular}
\caption{Finetuning data scaling law for 5\,m simulated fibre TMs using autoencoder with 124$\times$ compression. Experimental data used for early stopping and finetuning: per seed results.}
\label{tab:sm6-5m_exp_data_scaling_ES_FT_124x}
\end{center}
\end{table}

\begin{table}[htbp]
\begin{center}
\begin{tabular}{cccc} 
 \toprule
 & & \multicolumn{2}{c}{Validation error (\%)} \\ \cmidrule(lr){3-4}
 \shortstack{N experimental\\finetuning TMs} & Seed & \shortstack{Avg. finetuned error\\$\pm$ SD} & \shortstack{Avg. gain vs\\pretraining$\pm$ SD} \\
 \midrule
 0 & Avg. & $30.21 \pm 0.96$ & - \\
 & 42 & 30.94 & - \\
 & 43 & 29.12 & - \\
 & 44 & 30.58 & - \\
 5 & Avg. & $18.36 \pm 4.10$ & 11.85 \\
 & 42 & 23.06 & 7.88 \\
 & 43 & 15.53 & 13.59 \\
 & 44 & 16.50 & 14.08 \\
 10 & Avg. & $15.03 \pm 0.31$ & 15.18 \\
 & 42 & 15.32 & 15.62 \\
 & 43 & 14.71 & 14.41 \\
 & 44 & 15.06 & 15.52 \\
 163 & Avg. & $14.08 \pm 0.77$ & 16.14\\
 & 42 & 14.85 & 16.09 \\
 & 43 & 14.07 & 15.05 \\
 & 44 & 13.31 & 17.27 \\
 \bottomrule
\end{tabular}
\caption{Finetuning data scaling law for 5\,m simulated fibre TMs using autoencoder with 62$\times$ compression. Experimental data used for early stopping and finetuning: per seed results.}
\label{tab:sm7-5m_exp_data_scaling_ES_FT_62x}
\end{center}
\end{table}

\begin{table}[htbp]
\begin{center}
\begin{tabular}{lccccc}
 \toprule
 & \multicolumn{4}{c}{Average validation error $\pm$ SD (\%)} \\ \cmidrule(lr){3-6}
 \shortstack{Model type} & seed & \shortstack{N=0} & \shortstack{N=5} & \shortstack{N=10} & \shortstack{N=163}\\
 \midrule
 \shortstack{AE finetuned} & Avg. & $28.41 \pm 0.79$ & $14.79 \pm 0.58$ & $13.43 \pm 0.18$ & $11.59 \pm 0.23$ \\
 & 42 & 28.77 & 14.44 & 13.27 & 11.24 \\
 & 43 & 27.59 & 14.49 & 13.30 & 11.59 \\
 & 44 & 27.56 & 15.30 & 13.45 & 11.63 \\
 & 45 & 29.31 & 14.21 & 13.40 & 11.60 \\
 & 46 & 28.84 & 15.53 & 13.72 & 11.88 \\
 \lightrule
 \shortstack{AE, exp. only,\\all runs} & Avg. & - & $106.73 \pm 3.95$ & $106.88 \pm 7.18$ & $49.73 \pm 46.49$ \\
 & 42 & - & 103.24 & 101.84 & 16.16 \\
 & 43 & - & 106.50 & 103.26 & 17.29 \\
 & 44 & - & 108.28 & 108.54 & 13.93 \\
 & 45 & - & 112.59 & 118.75 & 101.20 \\
 & 46 & - & 103.05 & 101.99 & 100.08 \\
 \lightrule
 \shortstack{AE, exp. only,\\converged runs} & Avg. & - & none (0/5) & none (0/5) & $15.79 \pm 1.71$ (3/5) \\
 & 42 & - & - & - & 16.16 \\
 & 43 & - & - & - & 17.29 \\
 & 44 & - & - & - & 13.93 \\
 \lightrule
 PCA, sim. fit & - & 90.78 & - & - & - \\
 \lightrule
 PCA, exp. fit & - & - & 25.32$^{\dagger}$ & 1.68 & 0.77 \\
 \bottomrule
\end{tabular}
\caption{Autoencoder (AE) versus PCA at 62$\times$ compression on the original experimental TMs (validation set). $N$ is the number of experimental TMs used to fine-tune the simulation-pretrained AE ($N=0$: pretrained only, no fine-tuning), to train the AE from scratch (exp. only; ``converged'' runs have validation error below 90\%), or to fit PCA (exp. fit); PCA sim. fit uses 163 simulated TMs. AE entries are mean $\pm$ SD over five seeds and PCA entries are single deterministic fits. The AE checkpoint was chosen using the 20 validation TMs, so AE values are optimistic. Per seed results. $^{\dagger}$Rank-limited to 724 components.}
\label{tab:sm8-62x_PCA_vs_AE}
\end{center}
\end{table}

\begin{table}[htbp]
\begin{center}
\small
\setlength{\tabcolsep}{4pt}
\begin{tabular}{lcccccc}
 \toprule
 & \multicolumn{6}{c}{Validation error (\%)} \\ \cmidrule(lr){2-7}
 Model & seed 42 & seed 43 & seed 44 & seed 45 & seed 46 & Avg. $\pm$ SD\\
 \midrule
 \multicolumn{6}{l}{\textit{Raw second-fibre TMs}} \\
 AE sim-pretrained & 59.31 & 58.80 & 59.28 & 59.16 & 58.64 & 59.04 $\pm$ 0.30\\
 AE fine-tuned on 5 original TMs & 62.51 & 62.53 & 61.50 & 62.82 & 61.32 & 62.13 $\pm$ 0.68\\
 AE fine-tuned on 10 original TMs & 63.15 & 63.46 & 62.82 & 62.89 & 61.95 & 62.86 $\pm$ 0.57\\
 AE fine-tuned on 163 original TMs & 62.62 & 62.73 & 62.07 & 62.67 & 63.03 & 62.62 $\pm$ 0.35\\
 AE exp-only, 163 original TMs & 62.39 & 62.25 & 63.04 & 100.42 & 100.03 & 77.63 $\pm$ 20.63\\
 PCA fitted on 163 original TMs & \multicolumn{6}{c}{54.62 (single fit, no seeds)} \\
 PCA fitted on 163 simulated TMs & \multicolumn{6}{c}{92.43 (single fit, no seeds)} \\
 PCA fitted on second-fibre training TMs & \multicolumn{6}{c}{39.12 (single fit, no seeds)} \\
 \midrule
 \multicolumn{7}{l}{\textit{Band-limited second-fibre TMs (15 cycles/frame)}} \\
 AE sim-pretrained & 38.15 & 37.59 & 38.36 & 38.00 & 37.25 & 37.87 $\pm$ 0.45\\
 AE fine-tuned on 5 original TMs & 35.14 & 35.65 & 35.96 & 34.63 & 34.81 & 35.24 $\pm$ 0.56\\
 AE fine-tuned on 10 original TMs & 35.31 & 35.74 & 36.19 & 35.44 & 35.82 & 35.70 $\pm$ 0.34\\
 AE fine-tuned on 163 original TMs & 34.91 & 34.87 & 34.15 & 35.61 & 35.52 & 35.01 $\pm$ 0.59\\
 AE exp-only, 163 original TMs & 38.22 & 36.77 & 35.39 & 100.48 & 100.03 & 62.18 $\pm$ 34.77\\
 PCA fitted on 163 original TMs & \multicolumn{6}{c}{30.07 (single fit, no seeds)} \\
 PCA fitted on 163 simulated TMs & \multicolumn{6}{c}{91.15 (single fit, no seeds)} \\
 PCA fitted on second-fibre training TMs & \multicolumn{6}{c}{7.02 (single fit, no seeds)} \\
 \bottomrule
\end{tabular}
\caption{Zero-shot transfer to the second fibre: per seed results on the second-fibre validation TMs (62$\times$ compression). Nothing is refitted on the second-fibre data, except the in-domain PCA reference in the last row of each block. PCA rows are single deterministic fits, so one value is shown. AE = autoencoder; fine-tuning uses original-fibre TMs only.}
\label{tab:sm9-zhejiang_zero_shot}
\end{center}
\end{table}

\begin{table}[htbp]
\begin{center}
\small
\setlength{\tabcolsep}{4pt}
\begin{tabular}{lccccccc}
 \toprule
 & \multicolumn{5}{c}{AE validation error (\%) by seed} & & \\ \cmidrule(lr){2-6}
 $N$ & 42 & 43 & 44 & 45 & 46 & AE mean $\pm$ SD & PCA \\
 \midrule
 \multicolumn{8}{l}{\textit{Original TMs (50\,$\mu$m)}} \\
 0 & 28.77 & 27.59 & 27.56 & 29.31 & 28.84 & $28.41 \pm 0.79$ & - \\
 1 & 19.06 & 18.87 & 20.09 & 18.65 & 18.95 & $19.12 \pm 0.56$ & 89.50 \\
 2 & 14.33 & 15.25 & 15.19 & 15.27 & 15.98 & $15.31 \pm 0.42$ & 65.60 \\
 3 & 15.17 & 16.64 & 15.34 & 15.53 & 16.94 & $15.92 \pm 0.81$ & 52.40 \\
 5 & 14.53 & 14.87 & 15.23 & 15.35 & 16.08 & $15.21 \pm 0.58$ & 25.32 \\
 10 & 13.57 & 13.54 & 13.66 & 13.46 & 14.11 & $13.67 \pm 0.26$ & 1.68 \\
 all (163) & 11.56 & 11.59 & 11.56 & 11.73 & 11.73 & $11.64 \pm 0.09$ & 0.77 \\
 \midrule
 \multicolumn{8}{l}{\textit{Second fibre, raw}} \\
 0 & 59.31 & 58.80 & 59.28 & 59.16 & 58.64 & $59.04 \pm 0.30$ & - \\
 1 & 52.58 & 52.96 & 53.18 & 52.50 & 52.77 & $52.80 \pm 0.28$ & 94.17 \\
 2 & 52.37 & 52.31 & 52.66 & 51.98 & 52.45 & $52.35 \pm 0.25$ & 92.34 \\
 3 & 52.50 & 52.18 & 51.92 & 52.02 & 52.44 & $52.21 \pm 0.25$ & 86.08 \\
 5 & 51.95 & 52.14 & 51.36 & 51.85 & 51.97 & $51.86 \pm 0.30$ & 78.19 \\
 10 & 50.97 & 50.94 & 50.86 & 50.81 & 51.36 & $50.99 \pm 0.22$ & 60.77 \\
 all (131) & 50.18 & 50.23 & 50.14 & 50.16 & 50.10 & $50.16 \pm 0.05$ & 39.12 \\
 \midrule
 \multicolumn{8}{l}{\textit{Second fibre, band-limited}} \\
 0 & 38.15 & 37.59 & 38.36 & 38.00 & 37.25 & $37.87 \pm 0.45$ & - \\
 1 & 23.10 & 24.56 & 25.70 & 23.61 & 24.71 & $24.34 \pm 1.02$ & 91.30 \\
 2 & 21.77 & 23.08 & 21.46 & 21.53 & 21.22 & $21.81 \pm 0.73$ & 84.77 \\
 3 & 23.15 & 22.26 & 21.99 & 21.53 & 22.95 & $22.38 \pm 0.67$ & 74.34 \\
 5 & 22.45 & 22.31 & 20.95 & 21.51 & 20.79 & $21.60 \pm 0.76$ & 57.05 \\
 10 & 19.66 & 20.54 & 19.95 & 21.17 & 19.84 & $20.23 \pm 0.62$ & 14.59 \\
 all (131) & 18.44 & 18.87 & 18.66 & 18.74 & 18.97 & $18.73 \pm 0.20$ & 7.02 \\
 \bottomrule
\end{tabular}
\caption{Few-shot calibration: validation error (\%) of the autoencoder (AE) by seed, with its mean $\pm$ SD, and of PCA (a single fit). The simulation-pretrained checkpoint is fine-tuned on the first $N$ TMs of the target dataset and scored on that dataset's validation TMs; $N=0$ is the unadapted model, and ``all'' is every training TM of the dataset (163 original, 131 second fibre). PCA is fitted on the same $N$ TMs at 62$\times$ compression; for $N \le 5$ it is rank-limited. Fine-tuning uses a fresh AdamW optimiser, so values differ slightly from the fine-tuning table in the main text.}
\label{tab:sm10-zhejiang_few_shot}
\end{center}
\end{table}
\begin{table}[htbp]
\begin{center}
\small
\setlength{\tabcolsep}{4pt}
\begin{tabular}{lccccccc}
 \toprule
 & \multicolumn{5}{c}{AE validation error (\%) by seed} & & \\ \cmidrule(lr){2-6}
 $N+N$ & 42 & 43 & 44 & 45 & 46 & AE mean $\pm$ SD & PCA \\
 \midrule
 \multicolumn{8}{l}{\textit{Scored on original TMs (50\,$\mu$m)}} \\
 1+1 & 18.49 & 18.97 & 19.14 & 18.63 & 18.96 & $18.84 \pm 0.27$ & 82.49 \\
 2+2 & 16.80 & 17.52 & 17.32 & 16.97 & 18.74 & $17.47 \pm 0.76$ & 58.58 \\
 3+3 & 17.54 & 17.60 & 17.81 & 17.78 & 18.49 & $17.84 \pm 0.38$ & 42.14 \\
 5+5 & 16.89 & 17.16 & 17.11 & 17.43 & 17.54 & $17.22 \pm 0.26$ & 21.10 \\
 10+10 & 16.70 & 17.32 & 16.29 & 17.06 & 16.32 & $16.74 \pm 0.45$ & 27.30 \\
 131+131 & 15.64 & 15.27 & 15.64 & 15.62 & 15.47 & $15.53 \pm 0.16$ & 13.44 \\
 \midrule
 \multicolumn{8}{l}{\textit{Scored on second-fibre TMs (raw)}} \\
 1+1 & 56.86 & 56.71 & 57.09 & 56.53 & 56.84 & $56.81 \pm 0.20$ & 88.29 \\
 2+2 & 54.82 & 55.85 & 57.04 & 55.46 & 56.17 & $55.87 \pm 0.83$ & 80.65 \\
 3+3 & 54.21 & 54.50 & 55.93 & 54.16 & 55.08 & $54.77 \pm 0.74$ & 68.97 \\
 5+5 & 53.86 & 53.89 & 54.29 & 53.89 & 53.90 & $53.97 \pm 0.18$ & 54.87 \\
 10+10 & 53.47 & 53.68 & 53.38 & 53.45 & 53.03 & $53.40 \pm 0.24$ & 56.70 \\
 131+131 & 51.87 & 51.67 & 51.92 & 51.58 & 51.79 & $51.77 \pm 0.14$ & 38.30 \\
 \bottomrule
\end{tabular}
\caption{One shared model for both fibres: validation error (\%) by seed, with mean $\pm$ SD, for the autoencoder (AE) and for PCA (a single fit). The simulation-pretrained checkpoint is fine-tuned on the first $N$ original TMs and the first $N$ second-fibre TMs together ($2N$ TMs; $N=1$ uses two TMs), with the checkpoint chosen on the mean of the two validation errors, and is scored on each validation set. The shared PCA is one basis (62$\times$ compression) fitted on the same $2N$ TMs; it is rank-limited for small $N$. All uses 131 TMs from each fibre.}
\label{tab:sm11-shared_model}
\end{center}
\end{table}

\begin{figure}[htbp]
    \centering
    \includegraphics[width=1\textwidth]{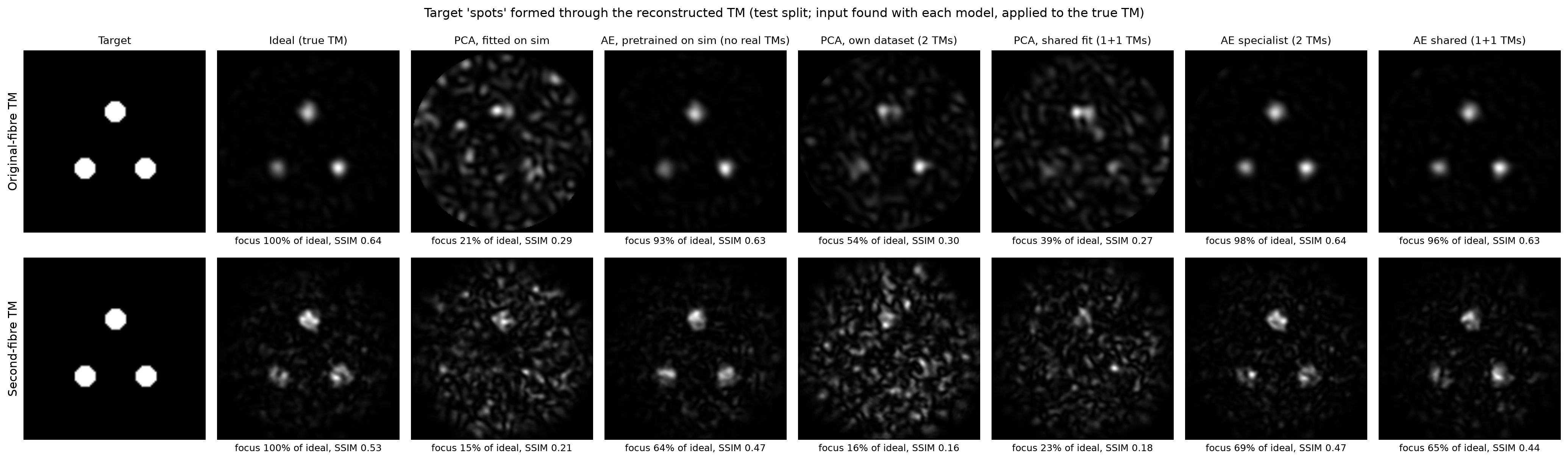}
    \caption{Focusing and imaging through reconstructed TMs for the target of spots (frozen test split, 62$\times$ compression), laid out as the main-text figure for the target ``F''. Rows: the first test TM of the original fibre and of the second fibre, the same for every model. Columns: target; ideal (the true TM, the best possible with 145 inputs); PCA fitted on 163 simulated TMs; autoencoder pretrained on simulation only; PCA fitted on two TMs of the fibre shown; PCA fitted on one TM from each fibre; autoencoder specialist (two TMs of the fibre shown); shared autoencoder (one TM from each fibre). Under each panel: mean focus enhancement as a percentage of the ideal and mean SSIM over all test TMs of that fibre (21 original, 17 second fibre).}
    \label{fig:sm-unscramble_spots}
\end{figure}

Supplementary Note S1: Band-limit radius. For each dataset we measured the diameter of the lit region as a fraction of the frame, $f$, and the Fourier radius $r_{99}$ enclosing 99\% of the field power, using 20 random speckle patterns from each of five validation TMs. The number of modes per speckle scales as $f^2 R^2$, so matching it between the datasets gives $R = r_{99}^{\mathrm{orig}} f_{\mathrm{orig}} / f_{\mathrm{sec\_fibre}}$. With $r_{99} = 16$, $f = 0.99$ (original) and $r_{99} = 37$, $f = 1.05$ (second fibre), this gives $R = 15$ cycles per frame. The Fourier disc holding 99\% of a speckle’s power contains about $\pi r_{99}^2 \approx 800$ complex degrees of freedom for the original data and about $4{,}300$ for the second-fibre data. This is an approximate mode count.

\end{document}